# M.A. Melkumov, I.A. Bufetov, K.S. Kravtsov, A.V. Shubin, E.M. Dianov.

# Absorption and emission cross section of $Yb^{3+}$ ions in $Al_2O_3$ and $P_2O_5$ doped fibers

*Abstract*

The results of measurements of the emission and absorption cross section for the transition $^2F_{5/2}$ -> $^2F_{7/2}$ of $Yb^{3+}$ ions in $Al_2O_3$ and $P_2O_5$ doped silica fibers are presented. Different techniques based on spectroscopic data, and more direct technique involving optical saturation of the transition are employed. The data obtained indicate an essential difference in cross section spectra of $Yb^{3+}$ ions in alumosilicate and phosphosilicate fibers

М.А.Мелькумов, И.А.Буфетов, К.С.Кравцов, А.В.Шубин, Е.М.Дианов.

# Сечения поглощения и вынужденного излучения ионов $Yb^{3+}$ в силикатных световодах, легированных $P_2O_5$ и $Al_2O_3$.

*Аннотация*

Представлены результаты измерений сечений поглощения и излучения для перехода $^2F_{5/2} \rightarrow {}^2F_{7/2}$ ионов $Yb^{3+}$ в кварцевых световодах, легированных $Al_2O_3$ и $P_2O_5$. Для измерений использовались как методы, основанные на спектроскопических данных, так и более прямые методы, использующие явление насыщения лазерного перехода.

*Abstract*

The results of measurements of the emission and absorption cross section for the transition $^2F_{5/2} \rightarrow {}^2F_{7/2}$ of $Yb^{3+}$ ions in $Al_2O_3$ и $P_2O_5$ doped silica fibers are presented. Different techniques based on spectroscopic data, and more direct technique involving optical saturation of the transition are employed.

С о д е р ж а н и е

# Введение

Волоконные световоды на основе плавленого кварца в большинстве случаев являются идеальной средой для ионов $Yb^{3+}$ при создании волоконных лазеров и усилителей на их основе. На иттербиевых световодах созданы эффективные волоконные усилители и лазеры с выходной мощностью от нескольких ватт до нескольких киловатт (см., напр., [1], [2], [3], [4], [5] и ссылки в этих работах). Создание таких устройств предполагает предварительное численное моделирование и оптимизацию их параметров. Но до настоящего времени в литературе имеется только очень ограниченная информация о сечениях лазерного перехода ионов иттербия в средах, характерных для волоконных световодов на основе плавленого кварца, причем в работах отсутствует достаточная информация о химическом составе световодов. Так, только в работах [1], [2] приведены спектральные зависимости сечений поглощения и вынужденного излучения ионов иттербия в германосиликатном стекле (без указания конкретного состава). В опубликованных работах по измерению параметров ионов $Yb^{3+}$ в объемных образцах [6], [7], [8], составы стекол сильно отличаются от состава сердцевины световодов исследуемых в настоящей работе. Поскольку во всех работах наблюдается существенная зависимость сечений переходов от состава стекла, опубликованные данные не могут быть использованы для моделирования волоконных лазеров. Кроме того, сравнение результатов, полученных на объемных образцах и в световодах, указывает на возможность не вполне корректного определения сечения вынужденного излучения ионов иттербия в объемных образцах из-за явления перепоглощения излучения.

В настоящее время, по-видимому, в большинстве случаев в волоконных лазерах используются два типа иттербиевых световодов: световоды на основе плавленого кварца с сердцевиной, легированной $P_2O_5$ (фосфоросиликатные (ФС) световоды), и с сердцевиной, легированной $Al_2O_3$ и небольшим количеством $GeO_2$ (алюмосиликатные (АС) световоды). Указанные легирующие добавки – $P_2O_5$, $Al_2O_3$ и $GeO_2$ необходимы для формирования профиля показателя преломления световода и для реализации однородного введения Yb в матрицу стекла, устранения явления кластеризации и снижения оптических потерь в световодах.

Целью данной работы явилось определение сечений поглощения и излучения перехода $^2F_{5/2}$ - $^2F_{7/2}$ ионов $Yb^{3+}$, введенных в сердцевину ФС и АС световодов. Был исследован ряд образцов ФС световодов, концентрация фосфора и иттербия в сердцевине которых изменялась в пределах 4÷10 и 1÷8 весовых %, соответственно. В АС световодах концентрация Al и Yb находилась в диапазоне 1÷2 и 1÷3 весовых %.

Заготовки световодов изготавливались по MCVD технологии. Введение иттербия и Al в АС световоды и иттербия в ФС световоды выполнялось как по растворной технологии (в одних образцах), так и из газовой фазы – в других. Проведенные измерения практически не выявили заметных отличий между световодами каждого типа (ФС или АС), независимо от концентрации легирующих добавок в указанных выше пределах и от способа введения некоторых из них (из раствора или из газовой фазы). Однако световоды различных типов (ФС и АС) показали существенно различные характеристики. Поэтому в дальнейшем изложении обычно указывается только тип световода, и не указываются более точно остальные его параметры.

## Схема уровней иона Yb³⁺

В этом разделе приведем для справок некоторые сведения о структуре энергетических уровней ионов иттербия в стекле.

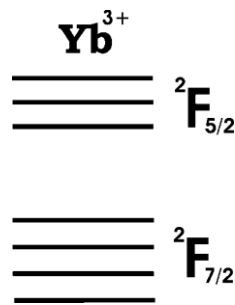

*Рис. 1 Структура уровней иона Yb³⁺ в стекле*

Ионы $Yb^{3+}$ в ближнем ИК и оптическом диапазоне имеют только один переход и, соответственно, только два энергетических уровня (Рис. 1). В приближении связи Рассела-Саундерса (LS-связь) эти уровни $Yb^{3+}$ обозначаются как $^2F_{5/2}$ и $^2F_{7/2}$. Полный момент этих уровней J равен 5/2 и 7/2, соответственно. Кратность вырождения уровней равна 2J+1 и составляет 6 и 8. Электрическое кристаллическое поле матрицы стекла снимает вырождение, но не полностью: поскольку ион $Yb^{3+}$ содержит нечетное число электронов, то, согласно теореме Крамерса, уровни при наложении любого внешнего электрического поля остаются двукратно вырожденными [9], разделяясь на 3 и 4 компоненты, соответственно. Полностью вырождение может быть снято в магнитном поле.

В матрице стекла вследствие неоднородного уширения спектры поглощения и люминесценции ионов иттербия представляют собой довольно широкие непрерывные линии в области 0.9 ÷ 1.1 мкм. Для описания процессов поглощения и вынужденного испускания квантов используют обычно понятия сечения поглощения и сечения вынужденного излучения. Для отдельных ионов с невырожденными энергетическими уровнями сечения поглощения и излучения равны, с вырожденными уровнями – пропорциональны.

Систему уровней ионов иттербия в стекле можно рассматривать как квазидвухуровневую при условии, что в каждом мультиплете устанавливается термодинамически равновесное распределение населенностей по подуровням. Но описание этой системы как двухуровневой приводит к тому, что сечения поглощения и вынужденного излучения оказываются существенно различными функциями длины волны [11] $\sigma_a(\lambda)$ и $\sigma_e(\lambda)$, которые, в свою очередь, зависят от вида термодинамического распределения ионов по подуровням, то есть от температуры. В настоящей работе в большинстве случаев речь идет о функциях $\sigma_a(\lambda)$ и $\sigma_e(\lambda)$ при комнатной температуре (Т ≈ 293°К), если температура не указана специально.

## Методы измерений

Существует большое количество методов измерения сечений лазерных переходов ионов $Re^{3+}$ в стеклах и в волоконных световодах (см напр. [10]). Методы, используемые при измерении сечений для одного типа редкоземельных ионов (например, $Er^{3+}$), часто пригодны для измерения сечений других типов редкоземельных ионов. По сравнению с объемными образцами, основным отличием при измерении сечений в световодах является неоднородность поля в сердцевине, а так же, как правило, неоднородное распределение активных ионов по радиусу световода. Это обстоятельство усложняет измерение сечений в световодах.

В данной работе измерение сечений проводилось несколькими независимыми методами, что увеличивает надежность полученных данных. Были независимо измерены как сечения поглощения, так и сечения вынужденного излучения, а так же из теории МакКамбера (D.E.McCumber, [11], далее ТМК) получено их отношение.

Для измерения сечения поглощения в световодах были использованы методы, описанные, например, в работах [10] и [12] и основанные на 1) измерении поглощения слабого сигнала в световоде (метод поглощения слабого сигнала), 2) на наблюдении насыщения люминесценции при увеличении мощности накачки (метод насыщения люминесценции), и 3) на наблюдении насыщения поглощения в отрезке световода при увеличении мощности накачки (обозначим его как метод поглощения большого сигнала).

Различные методы предъявляют различные требования к характеру необходимых измерений. Так, для реализации метода поглощения слабого сигнала необходимо знать абсолютное значение распределения ионов $Yb^{3+}$ по радиусу сердцевины световода и профиль распределения интенсивности излучения по поперечному сечению моды. Для метода поглощения большого сигнала (и метода насыщения люминесценции) необходимо знать распределение интенсивности излучения в световоде по радиусу, относительный профиль распределения ионов $Yb^{3+}$ по поперечному сечению, время жизни ионов на верхнем лазерном уровне и зависимости абсолютных значений мощности на выходе из световода от мощности на его входе (для метода поглощения большого сигнала) или зависимости относительной интенсивности люминесценции от мощности накачки в световоде (для метода насыщения люминесценции. Существенно, что для двух данных методов не нужно знать абсолютную концентрацию ионов $Yb^{3+}$.

Сечение вынужденного излучения в настоящей работе определялось по спектрам люминесценции и времени жизни ионов $Yb^{3+}$ на верхнем лазерном уровне при помощи ТМК [11]. Кроме того, ТМК позволяет установить однозначное соответствие между сечениями излучения и поглощения ([11], [13]), что дает возможность, например, получить сечение поглощения из сечения вынужденного излучения. Но для этого необходимо знать положение энергетических уровней в мультиплетах. Расчеты сечения вынужденного излучения по спектру люминесценции при помощи ТМК не требуют знания концентрации ионов $Yb^{3+}$ и профиля интенсивности в световоде. В настоящей работе, также, рассматриваются основные допущения ТМК и область ее применимости.

## Определение сечения поглощения по поглощению слабого сигнала.

Метод измерения сечения поглощения по поглощению слабого сигнала является наиболее простым для определения сечений [10], однако он обладает тем недостатком, что требует знания абсолютной концентрации ионов $Yb^{3+}$. Данное обстоятельство может существенно снижать его точность, т.к. измерение концентрации ионов $Yb^{3+}$ в световоде представляет собой достаточно сложную задачу.

В простейшей ситуации, когда интенсивность и состав образца стекла не зависят от поперечных координат (одномерный случай), сечение поглощения $\sigma_a$ может быть найдено из коэффициента поглощения слабого сигнала $\alpha$ (см$^{-1}$), и концентрации ионов $Yb^{3+}$ $n$ (см$^{-3}$) как $\sigma_a = \alpha/n$. В случае волоконного световода ситуация осложняется тем, что как интенсивность излучения, так и концентрация ионов существенно зависят от радиуса. Для одномодовых световодов (в настоящей работе рассматриваются только такие волоконные световоды) расчет можно провести следующим способом. Обозначим через $I(r)$ интенсивность излучения по поперечному сечению, через $z$ - продольную координату, а через $P$ −полную мощность излучения, проходящую по световоду. Тогда если $n(r)$ – концентрация ионов $Yb^{3+}$ в стекле, то

$$\frac{dP}{dz} = -2\pi\,\sigma_a \int_0^\infty n(r) I(r,z) r\,dr, \qquad (1)$$

и

$$\alpha = -\frac{1}{P}\frac{dP}{dz} = \sigma_a \frac{\int\limits_0^\infty n(r)I(r)rdr}{\int\limits_0^\infty I(r)rdr} = \sigma_a n_{eff} \qquad (2)$$

Таким образом, для нахождения искомого сечения поглощения необходимо вычислить $n_{eff}$ —эффективную концентрацию, учитывающую распределение поля моды:

$$n_{eff} = \frac{\int\limits_0^\infty n(r)I(r)rdr}{\int\limits_0^\infty I(r)rdr}, \qquad (3)$$

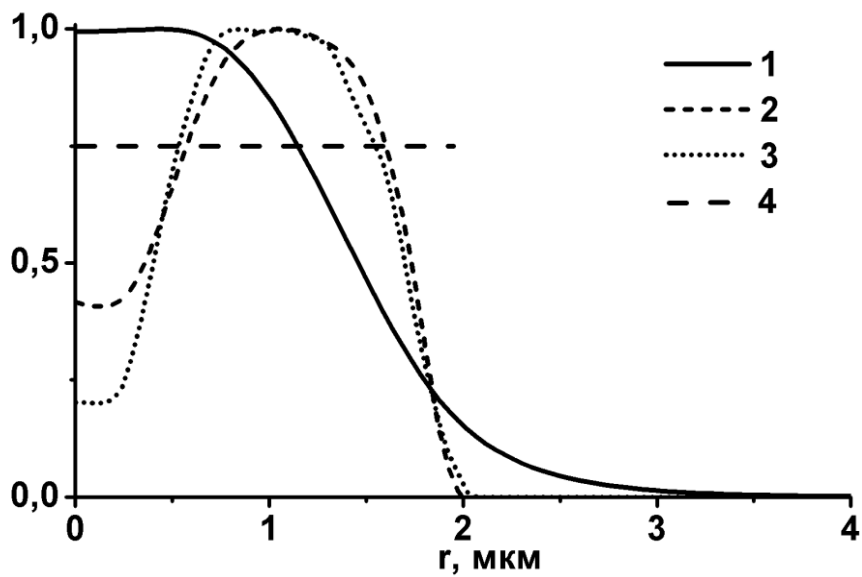

*Рис. 2. Распределение поля моды -1, показателя преломления -2 и концентрации ионов иттербия-3 по поперечному сечению световода (все нормировано на единицу), 4 – уровень $n_{eff}$.*

На Рис. 2 представлены типичные зависимости распределения интенсивности, профиля показателя преломления и концентрация ионов Yb$^{3+}$ в световоде по радиусу волоконного световода. Распределение интенсивности поля моды вычислялось по профилю показателя преломления согласно методике, описанной в работе [14].

Путем численного интегрирования (3) можно найти значение $n_{eff}$ и произвести расчет сечения поглощения перехода $^2F_{5/2} \to {}^2F_{7/2}$ ионов Yb$^{3+}$ для данной длины волны.

## *Определение сечения поглощения по поглощению большого сигнала*

Этот метод позволяет определить сечение поглощения по измеренной мощности насыщения [10]. Метод основан на том, что при распространении сигнала большой мощности (порядка мощности насыщения) в световоде наблюдается просветление среды, и, следовательно, меняются условия поглощения. Сначала рассмотрим одномерную картину (без зависимости от радиуса). Представим себе, что по световоду идет излучение накачки на такой длине волны, что сечение вынужденного излучения там пренебрежимо мало (много меньше сечения поглощения). Тогда, если $s$ – плотность потока фотонов накачки [1/(см$^2$·с)], то

$$\frac{ds}{dz} = -n_1 \sigma_a s,$$

где $n_1$-населенность нижнего уровня. При условии малости числа актов вынужденных переходов с верхнего уровня в основное состояние (т.е. пренебрегая усиленной люминесценцией и предполагая отсутствие генерации) должно выполняться равенство числа актов спонтанных переходов сверху вниз и вынужденных снизу вверх. Таким образом, можно записать

$$\frac{dn_2}{dt} = -\frac{n_2}{\tau} + n_1 \sigma_a s = -\frac{n_2}{\tau} + (n_0 - n_2)\sigma_a s = 0, \qquad (4)$$

где $n_2$-населенность верхнего уровня, $n_0$-полная концентрация ионов $Yb^{3+}$. Отсюда для $n_1$ и $n_2$ можно записать:

$$n_2 = n_0 \frac{\sigma_a s \tau}{\sigma_a s \tau + 1}, \qquad (5)$$

$$n_1 = n_0 \frac{1}{\sigma_a s \tau + 1}, \qquad (6)$$

откуда

$$\frac{ds}{dz} = -\frac{\sigma_a n_0 s}{1 + \sigma_a \tau s}, \qquad (7)$$

Как известно, данное дифференциальное уравнение имеет решение:

$$\ln\left(\frac{s(z)}{s(0)}\right) + \frac{1}{s_0}\left(s(z) - s(0)\right) = -\sigma_a n_0 z, \qquad (8)$$

где $s_0$ – поток квантов насыщения, определяемый как $s_0 = 1/\sigma_a \tau$.

Таким образом, найдена зависимость мощность накачки от координаты. В эксперименте удобнее измерять зависимость выходной мощности накачки от входной мощности при постоянной длине световода. Очевидно, эти две зависимости тесно связаны и могут быть пересчитаны одна в другую. В полученное уравнение входят всего два параметра: $n_0 \sigma_a$ и $s_0$, причем первый из них есть не что иное, как коэффициент поглощения, который непосредственно может быть измерен, а второй параметр $s_0$ должен быть выбран таким, чтобы расчетная зависимость совпадала с экспериментальными данными. Зная $s_0$ и время жизни $\tau$, получим искомое значение сечения поглощения $\sigma_a$. Ситуация осложняется тем, что данная модель одномерна, а в реальном световоде присутствует зависимость параметров от радиуса.

В настоящей работе проводился численный расчет, учитывающий зависимости интенсивности излучения накачки и концентрации ионов $Yb^{3+}$ от радиуса.

Пусть $P(z)$ – мощность накачки, распространяющейся по волоконному световоду, тогда

$$s(r) = \frac{P}{\hbar\omega} \frac{I(r)}{2\pi \int I(r)rdr}, \qquad (9)$$

где $I(r)$ – распределение интенсивности по радиусу в моде световода, $\hbar$ - постоянная Планка, $\omega$ -частота излучения накачки. Зная $s(r)$, вычисляем производную от мощности $P$ по координате

$$\frac{dP}{dz} = -2\pi\hbar\omega \int r \frac{\sigma_a n_0(r) s(r)}{1+\sigma_a \tau s(r)} dr, \qquad (10)$$

далее выразим произведение $\sigma_a n_0(r)$ через коэффициент поглощения $\alpha$. Для этого для начала запишем:

$$\sigma_a n_0(r) = \sigma_a n(0) n_{norm}(r), \qquad (11)$$

где $n(0)$- концентрация ионов $Yb^{3+}$ на оси световода, $n_{norm}(r)$ -профиль концентрации нормированный на единицу на оси световода. Из формулы (2) и (11) имеем:

$$\alpha = \sigma_a n(0) \frac{\int_0^\infty n_{norm}(r) I(r) rdr}{\int_0^\infty I(r) rdr}, \qquad (12)$$

отсюда для произведения $\sigma_a n_0(r)$ получим:

$$\sigma_a n_0(r) = \frac{\alpha}{C} n_{norm}(r),$$

где

$$C = \frac{\int_0^\infty n_{norm}(r) I(r) r\, dr}{\int_0^\infty I(r) r\, dr}$$

Введем обозначение:

$$const = \frac{2\pi hc}{\lambda} \int I(r) r\, dr,$$

и перепишем выражение (10) в более удобном для расчетов виде:

$$\frac{dP(z)}{dz} = -\frac{2\pi hc}{\lambda} \frac{\alpha}{C} \int \frac{n_{norm}(r)\, r\, P(z)\, I(r)}{const + \sigma_a \tau P(z) I(r)} dr, \qquad (13)$$

Данная формула позволяет численно рассчитать зависимость изменения мощности в световоде от координаты $P(z)$ для данного значения сечения $\sigma_a$. Из этой зависимости легко получить зависимость мощности на выходе световода определенной длины от мощности на входе. Такая же зависимость измеряется в ходе эксперимента. В итоге, необходимо подобрать такое значение $\sigma_a$, при котором расчетная кривая будет максимально совпадать с экспериментальной зависимостью. При этом будет однозначно найдено искомое значение сечения поглощения $\sigma_a$.

Как видно из формулы (13), для определения сечения поглощения данным методом требуется только относительное распределение концентрации ионов $Yb^{3+}$ по сечению световода, время жизни и коэффициент поглощения.

Недостатками данного метода являются - во-первых – необходимость постоянного контроля отсутствия лазерной генерации, так как для измерений используются световоды длинной порядка нескольких сантиметров, что необходимо для обеспечения достаточной точности измерений, а входные мощности имеют порядок мощности насыщения и более, т.е. в световодах присутствует высокая инверсия населенности, что приводит к большим коэффициентам усиления. Во-вторых, данный метод весьма чувствителен к ошибкам в измерении входной мощности, то есть к потерям в точке сварки исследуемого световода со световодом накачки.

## *Определение сечения поглощения по насыщению люминесценции*

Данный метод часто используется для определения сечения поглощения в волоконных световодах [2], [10], [12]. Он позволяет получить значение сечения поглощения для определенного значения длины волны. Основное достоинство данного метода, так же как и у предыдущего метода, состоит в том, что он не требует знания абсолютных значений концентрации активных ионов в световоде. Суть метода состоит в следующем: при прохождении накачки по активному волоконному световоду ионы $Yb^{3+}$ переходят в возбужденное состояние и за счет спонтанных распадов люминесцируют. Измеряя интенсивность люминесценции от мощности накачки, можно вычислить искомое значение сечения поглощения (на длине волны накачки).

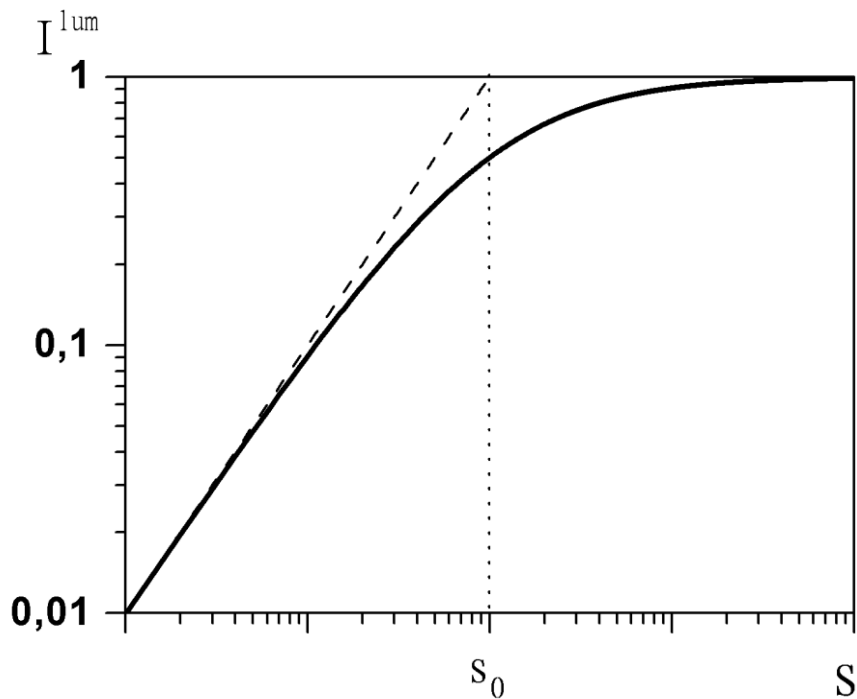

*Рис. 3. Интенсивность люминесценции в зависимости от интенсивности накачки*

Как уже было ранее установлено при условиях, когда можно пренебречь вынужденными переходами с верхнего уровня в основное состояние,

населенность верхнего лазерного уровня $n_2$ может быть выражена в виде (см. формулу (5)):

$$n_2 = n_0 \frac{\sigma_a \tau s}{1 + \sigma_a \tau s} \qquad (14)$$

С другой стороны, интенсивность спонтанной люминесценции пропорциональна населенности верхнего уровня, т.е. $n_2$. Рассмотрим теперь график зависимости $n_2$ от $s$ (см. Рис. 3). При малых интенсивностях накачки $n_2$ линейно растет, а при больших доходит до максимального значения (насыщается). Из простейших вычислений следует, что точка пересечения касательной к графику в нуле с уровнем насыщения люминесценции соответствует интенсивности накачки $s_0 = 1/\sigma_a \tau$ (поток квантов насыщения). Измерив эту зависимость и рассчитав $s_0$, по известному $\tau$ можно найти $\sigma_a$.

При наличии зависимости концентрации ионов $Yb^{3+}$ и интенсивности излучения от радиуса (в случае световода), расчеты несколько усложняются. В таком случае, необходимо вычислить населенности верхнего уровня $n_2(r)$ (см. уравнения (9), (14)) в каждой точке по радиусу, и затем проинтегрировать:

$$I^{lum} \sim \int_0^\infty n_2(r) r\, dr \sim \int_0^\infty \frac{\sigma_a \tau\, P\, I(r)\, n_{norm}(r)\, r}{const + \sigma_a \tau\, P\, I(r)} dr, \qquad (15)$$

где $I^{lum}$ – интенсивность люминесценции, остальные обозначения такие же, как в предыдущем параграфе. Далее проводился подбор значения сечения поглощения $\sigma_a$ до максимального соответствия данных численного расчета экспериментальной кривой. Отметим, что подбор значения сечения поглощения $\sigma_a$ в данном методе несколько отличается от алгоритма, рассмотренного в методе поглощения большого сигнала. Основное отличие состоит в том, что в данном методе измеряется относительное изменение интенсивности люминесценции от проходящей мощности, в то время как в предыдущем методе в эксперименте измерялись абсолютные значения мощности прошедшего и падающего сигналов. Таким образом, в данном методе необходима нормировка экспериментальных данных по интенсивности. Такая нормировка проводилась следующим образом: кривая, полученная по формуле (15), всегда нормировалась на единицу при мощности, стремящейся к бесконечности. Экспериментальные данные нормировались по оси ординат для каждого значения $\sigma_a$ с тем, чтобы иметь минимальное расхождение с расчетной кривой. Затем выбиралось новое значение $\sigma_a$, при котором расхождение с экспериментальными данными минимально, и нормировка по оси ординат повторялась. Таким образом, после нескольких итераций значение сечения становилось постоянным и практически не менялось с каждой следующей итерацией.

К достоинствам данного метода можно отнести то, что при определении сечения поглощения по данному методу не требуется знание абсолютной концентрации ионов иттербия в световоде и коэффициента поглощения. Так же данный метод принципиально не чувствителен к потерям на сварке световода накачки и исследуемого образца. Все это, несомненно, увеличивает достоверность результатов полученных данным методом.

### *Определение сечения вынужденного излучения по спектрам люминесценции и времени жизни*

По спектру люминесценции и времени жизни ионов в возбужденном состоянии можно рассчитать абсолютное значение сечения вынужденного излучения при помощи ТМК (D.E. McCumber [11], [13] ). Преимуществом данного способа определения сечения является то, что при расчетах не требуются какие-либо данные по концентрации ионов $Yb^{3+}$ в световоде, и не используется профиль распределения интенсивности излучения в световоде. Необходимыми экспериментальными данными для расчетов сечения вынужденного излучения с использованием ТМК являются время жизни $\tau$ на верхнем лазерном уровне и форма спектра эмиссии. Существенно, что ТМК позволяет рассчитать сечение поглощения в длинноволновой области ($\lambda$>1 мкм, где коэффициент поглощения мал), зная в этой области сечение вынужденного излучения. Благодаря этому, такой подход используется во многих работах по определению сечений в $Er^{3+}$ и $Yb^{3+}$ стеклах и световодах [13], [15], [16].

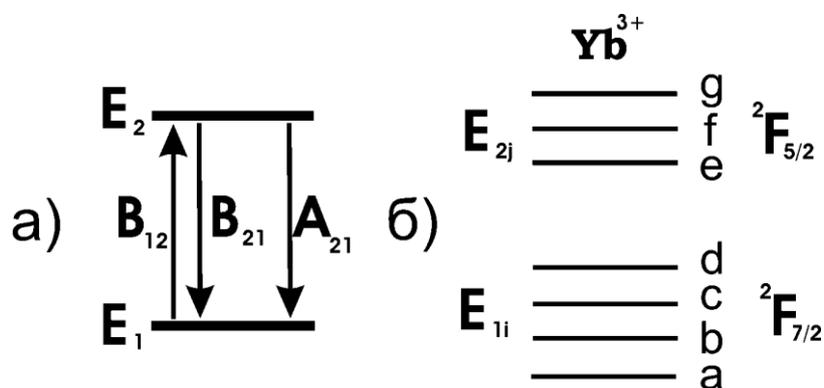

*Рис. 4. а) двухуровневая система, б) система уровней $Yb^{3+}$.*

На Рис. 4 представлены соответственно обычная двухуровневая схема, используемая при расчете вероятностей переходов по Эйнштейну – а), и двухуровневая схема с расщеплением на подуровни, отражающим структуру уровней иона $Yb^{3+}$ -б).

Рассмотрим простейшую теорию, основанную на чисто двухуровневой схеме. Коэффициенты Эйнштейна, связывающие вероятности вынужденных и спонтанных переходов, связаны следующими отношениями [11]:

$$A_{21} = 16\pi^2 \hbar \left(\frac{\omega}{2\pi c}\right)^3 B_{21},$$

$$g_1 B_{12} = g_2 B_{21},$$

где $g_i$ – коэффициент вырождения $i$-го уровня, $A_{21}$ – вероятность спонтанного перехода, а вероятность вынужденного перехода с i-го на j-й уровень задается выражением $B_{ij}\rho(v)$, где $\rho(v)$ – плотность энергии фотонов, $\omega$ – круговая частота фотона, участвующего во взаимодействии.

Аналогичные зависимости можно получить и для более общего случая, когда энергетические уровни расщеплены, и вся система обладает ненулевой температурой. В таком приближении, модель должна хорошо описывать ионы $Yb^{3+}$ в кварцевом стекле, так как у $Yb^{3+}$ нет промежуточных энергетических уровней, а значит, система максимально схожа с рассматриваемой моделью.

Рассмотрим такую систему, находящуюся в тепловом равновесии при температуре $T$. Обозначим через $E_{1i}$ уровни энергии примесных атомов в основном состоянии и $E_{2i}$ – в первом возбужденном состоянии, где $i$ – индекс, пробегающий по всем подуровням. Тогда в условиях теплового равновесия вероятность заселения каждого из подуровней равна

$$w_{mi} = \frac{\exp(-E_{mi}/kT)}{\sum_j \sum_b \exp(-E_{jb}/kT)}$$

где $w_{mi}$ – вероятность заселения уровня $E_{mi}$, $k$ – постоянная Больцмана.

Рассмотрим взаимодействие такой системы с фотонами частоты $\omega$. Вероятности переходов с какого-либо подуровня основного состояния - $(1i)$ на определенный подуровень возбужденного состояния - $(2j)$ описываются матрицей с элементами $M_{1i,2j}$. Каждый элемент данной матрицы определяет вероятность перехода с одного подуровня основного состояния на определенный подуровень возбужденного состояния. Причем для двух конкретных подуровней вероятности переходов сверху вниз и снизу вверх равны: $|M_{2j,1i}|^2 = |M_{1i,2j}|^2$.

Если обозначить концентрации возбужденных (на верхний уровень) и невозбужденных ионов иттербия, как $N_2$ и $N_1$ соответственно, то согласно определениям сечений излучения и поглощения, можно записать [11]

$$\frac{N_1 \sigma_a(\mathbb{k},\omega)_{12}}{N_2 \sigma_e(\mathbb{k},\omega)_{21}} = \frac{\sum\limits_{i,j} w_{1i} \left| M_{2j,1i}(\mathbb{k},\omega) \right|^2 \delta(E_{2j} - E_{1i} - \hbar\omega)}{\sum\limits_{i,j} w_{2j} \left| M_{1i,2j}(\mathbb{k},\omega) \right|^2 \delta(E_{2j} - E_{1i} - \hbar\omega)} \quad (16)$$

где $\mathbb{k}$ - волновой вектор. Но при тепловом равновесии

$$w_{1j} \delta(E_{2j} - E_{1i} - \hbar\omega) = w_{2i} \exp(\hbar\omega/kT) \delta(E_{2j} - E_{1i} - \hbar\omega) \quad (17)$$

Комбинируя (17) и (16), получим следующий результат:

$$\sigma_e(\mathbb{k},\omega)_{21} = \sigma_a(\mathbb{k},\omega)_{12} \frac{N_1}{N_2} \exp(-\hbar\omega/kT) = \sigma_a(\mathbb{k},\omega)_{12} \exp((\varepsilon - \hbar\omega)/kT)$$

$$(18)$$

Таким образом, в ТМК устанавливается однозначное соответствие между сечением вынужденного излучения и сечением поглощения для данной длины волны при данной температуре. Основным параметром, определяющим их соотношение, является фактор $N_1/N_2$ или $\varepsilon$. Причем, для отношения $N_1/N_2$ можно записать

$$\frac{N_1}{N_2} = e^{\varepsilon/kT} = \frac{\sum\limits_j \exp(-\frac{E_{1i}}{kT})}{\sum\limits_i \exp(-\frac{E_{2j}}{kT})} \quad (19)$$

С другой стороны, так же как и соотношения Эйнштейна для двухуровневой системы, в ТМК установлена связь между времени жизни на верхнем уровне τ и сечением эмиссии σ$_e$. При полном термодинамическом равновесии (т.е. когда вещество находится в тепловом равновесии с излучением), число вынужденных переходов с нижнего уровня в верхнее состояние должно равняться сумме вынужденных и спонтанных переходов сверху вниз. Исходя их этого факта, было найдено соотношение между временем жизни в возбужденном состоянии и сечением вынужденного излучения (см. напр. в [13]):

$$\frac{1}{\tau_{21}} = 8\pi c n^2 \int_0^\infty \frac{\sigma_e(\lambda)}{\lambda^4} d\lambda, \quad (20)$$

Зная время жизни и форму сечения вынужденного излучения (например, из спектра эмиссии) и, используя соотношение (20), легко получить абсолютное значение сечения люминесценции:

$$\sigma_e(\lambda) = \frac{\lambda^5}{8\pi c n^2 \tau} \frac{I^{lum}(\lambda)}{\int_0^\infty \lambda I^{lum}(\lambda) d\lambda}, \quad (21)$$

Заметим, что сечение люминесценции $\sigma_e(\lambda)$ и интенсивность спонтанной люминесценции $I^{lum}(\lambda)$ связаны следующим соотношением: $\sigma_e(\lambda) \sim \lambda^5 I^{lum}(\lambda)$ (см. приложение 2).

Далее по известной структуре уровней $Yb^{3+}$ в стекле (структуру уровней можно получить, например, по спектрам поглощения и люминесценции, снятым при низкой температуре) легко рассчитать величину отношения, $N_1/N_2$ а затем вычислить значение сечения поглощения $\sigma_a(\lambda)$.

### *Область применимости теории МакКамбера*

При выводе соотношения (18) принимается во внимание различная заселенность штарковских подуровней основного и возбужденного состояний. Однако, как видно из выражений (16) и (17), переходы между штарковскими подуровнями описаны дельта функциями - $\delta(E_{2j} - E_{1i} - \hbar\omega)$, что указывает на то, что ширины каждого из подуровней должны быть много меньше величины $kT$. Данное условие далеко не всегда выполняется в иттербиевых стеклах, ввиду того, что даже при комнатной температуре ($kT$~200 см$^{-1}$) неоднородное уширение линии (40÷300 см$^{-1}$) сопоставимо с величиной $kT$. В работе [17] рассмотрено влияние неоднородного и однородного уширения линии на результаты расчетов формы спектра сечения поглощения и излучения по ТМК. Поскольку вопросы применимости ТМК важны для результатов настоящей работы, то ниже приведены основные выводы, полученные в работе [17] относительно применимости развитого в ТМК подхода.

Как хорошо известно, неоднородное уширение линии перехода в твердом теле связано с неоднородностью вещества, что приводит к различному положению штарковских подуровней отдельных ионов. При этом суммарная форма линии переходов для всех ионов обычно описывается функцией Гаусса. Однородное уширение вызвано динамическим возмущением энергетических уровней из-за колебаний решетки и одинаково для всех ионов. В этом случае форма линии перехода для каждого атома описывается функцией Лоренца. Суммарная форма линии перехода представляет собой свертку функций Гаусса и Лоренца.

Перемножение спектра указанной формы на фактор $\exp((\varepsilon-\hbar\omega)/kT)$ при выполнении преобразования согласно ТМК, приводит к искажению формы спектра. Так, однородно уширенная линия, описываемая функцией Лоренца, спадает на краях по степенному закону, что приводит к существенным искажениям на краях при умножении на экспоненту. При этом длинноволновая часть полученного спектра оказывается завышенной, в то время как коротковолновое плечо, наоборот, меньше реально наблюдаемой формы спектра. В случае неоднородного уширения края спектра спадают по экспоненциальному закону, и фактор $\exp((\varepsilon-\hbar\omega)/kT)$ не вносит существенных искажений в крылья спектра.

Таким образом, искажения линии гауссовой формы носит более локальный характер, в то время как линия, описываемая функцией Лоренца, получает значительные искажения именно на краях [17]. Искажения спектра при однородном уширении становятся еще значительнее по сравнению с неоднородным уширением, для спектра, состоящего из нескольких линии, так как в этом случае ошибки на краях суммируются. Поскольку основное условие для применимости ТМК имеет вид, $\delta E_{ij} \ll kT$, очевидно, что искажения, вносимые фактором $\exp((\varepsilon - \hbar\omega)/kT)$, будут зависеть от температуры $T$. При этом, следует учесть, что неоднородная ширина линии переходов редкоземельных ионов в стеклах практически не изменяется с увеличением температуры, в то время как однородная ширина возрастает как $T^2$ [18].

На Рис. 5 представлены зависимости от температуры относительного среднеквадратичного (СК) отклонения спектра, полученного при помощи ТМК, от истинного спектра для однородно и неоднородно уширенных линий различной ширины (из работы [17]). Для неоднородно уширенной линии ширина спектра остается постоянной во всем диапазоне температур, в то время как для однородно уширенной линии ширина спектра зависит от температуры. На графике приведено значение ширины линии при комнатной температуре.

Из рисунка видно, что среднеквадратичное отклонение для чисто неоднородно уширенной линии спадает по закону *1/T*. По мере уменьшения температуры ниже комнатной, роль искажения спектра вследствие неоднородного уширения становится доминирующей. Какой из механизмов – однородное или неоднородное уширение будет играть главную роль в искажении спектра при данной температуре зависит от конкретного типа стекла.

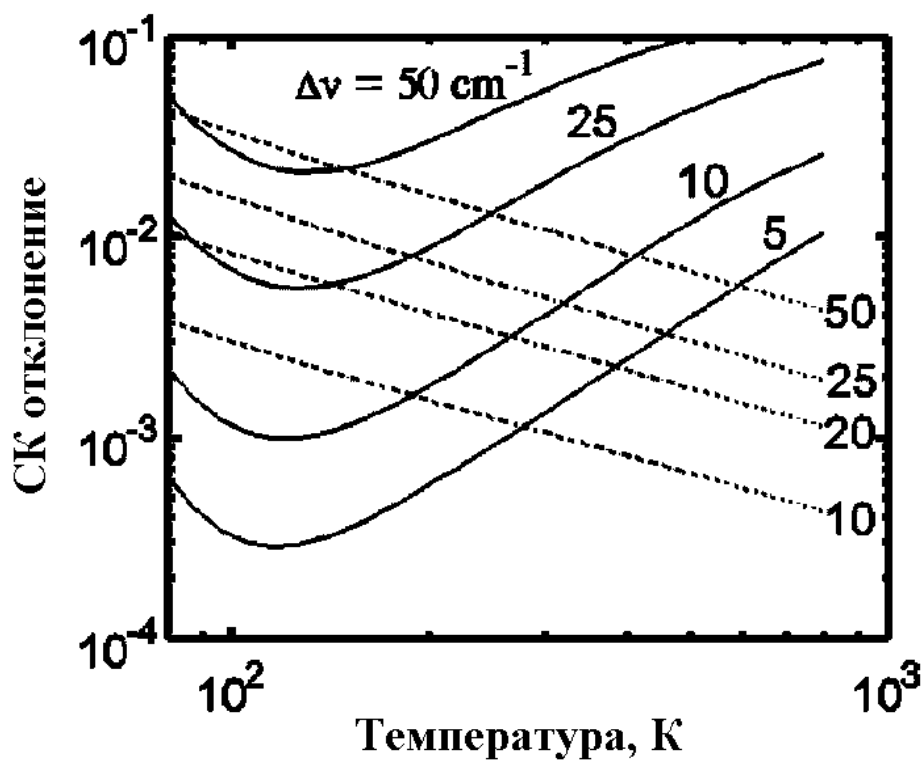

*Рис. 5 Температурная зависимость относительного среднеквадратичного-СК отклонения для чисто однородно (сплошные линии) или чисто неоднородно (пунктирные линии) уширенной линии [17], $\Delta v[см^{-1}]$-соответствующая ширина линий по полувысоте при комнатной температуре.*

В заключение данного раздела стоит отметить, что расчет абсолютного значения сечения вынужденного излучения по спектру люминесценции и времени жизни может быть выполнен не только на основе ТМК (см. формулу (21)), но и по формуле Фухтбауера-Ладенбурга (**Fuchtbauer-Ladenburg (FL)**) [10], [12]. Выражение, используемое для расчета сечения вынужденного излучения в **FL** методе, может быть получено через коэффициенты Эйнштейна *A* и *B*. Данный подход применим к переходам, для которых выполняется хотя бы одно из указанных ниже условий [19]: все штарковские подуровни обоих мультиплетов должны быть одинаково заселены ($\Delta E_{1,2} \ll kT$), или все переходы между любыми подуровнями имеют одинаковую вероятность.

Фактически каждое из указанных условий приводит к ситуации, идентичной вырожденному случаю (без расщепления) для которой применимы соотношения Эйнштейна. Для вычисления сечения вынужденного излучения в **FL** методе используется соотношение [12]

$$\sigma_e(\lambda) = \frac{<\lambda>^4}{8\pi c n^2 \tau} \frac{I^{lum}(\lambda)}{\int_0^\infty I^{lum}(\lambda) d\lambda}, \qquad (22)$$

здесь $<\lambda>$ - длина волны в максимуме линии излучения. Видно, что формулы для расчета сечения вынужденного излучения по спектрам люминесценции и времени жизни при помощи ТМК (21) и **FL** метода очень близки, поэтому, при сравнении полученных нами данных с результатами работ других авторов, мы будем цитировать результаты, полученные как с помощью формулы Фухтбауера-Ладенбурга, так и на основе ТМК.

## Экспериментальные результаты.

### Измерение концентрации $Yb^{3+}$ и профиля показателя преломления

Распределение концентрации ионов $Yb^{3+}$ и профиль показателя преломления по радиусу измерялись в заготовках. Пересчет их для приведения к волоконным световодам производился в предположении, что при вытяжке не происходит диффузия между слоями. Концентрация измерялась методом рентгеноспектрального микроанализа с помощью сканирующего электронного микроскопа JSM-5910LV (JEOL) и рентгеновского спектрометра INCA (Oxford Instruments). Концентрация атомов Yb определялась по двум спектральным линиям иттербия – $L_{\alpha 1,2}$ (≈7400 эВ) и $M_{\alpha 1,2}$ (≈1520 эВ). Поскольку линия $M_{\alpha 1,2}$ обладает меньшей энергией, она дает лучшее пространственное разрешение и, с этой точки зрения предпочтительнее. Однако в АС световодах возможно увеличение ошибки в определении концентрации Yb из-за присутствия близко расположенных линий алюминия ($\approx$ 1550 эВ). Как в АС, так и в ФС световодах результаты измерений концентрации иттербия по линиям $L_{\alpha 1,2}$ и $M_{\alpha 1,2}$ не совпадают, причем отличие достигает 50% и более. Поэтому нами проводились измерения концентрации иттербия по обеим линиям, дальнейший контроль точности измерений проводился при сравнении с данными, полученными другими методами (см. ниже).

Измерения профиля показателя преломления в заготовках проводились на установке P102 (York Technology).

### *Измерение спектров люминесценции*

Измерения спектров люминесценции световодов проводились с помощью спектроанализатора HP 71451B. В одномодовый световод вводилось излучение лазера на кристалле Ti:Al$_2$O$_3$ с длиной волны около 948 нм. Спектры измерялись в направлении, перпендикулярном оси активных световодов, что исключало их искажение вследствие перепоглощения. В качестве примера на Рис. 6 представлены спектры люминесценции в образцах АС и ФС световодов.

Следует отметить, что методика измерений спектров люминесценции перпендикулярно оси одномодовых оптических световодов выгодно отличается от подобных измерений с использованием объемных образцов. Так, в случае одномодовых световодов перепоглощение практически полностью исключено (диаметр сердцевины ~ 10 мкм), в то время как в объемных образцах перепоглощение может существенно искажать спектр люминесценции. В подтверждение данного факта отметим, что отношение пиков на 975 нм и на 1020 нм в спектрах люминесценции наших световодов составляет 4÷5 (примерно такое же отношение наблюдалось в [2]).

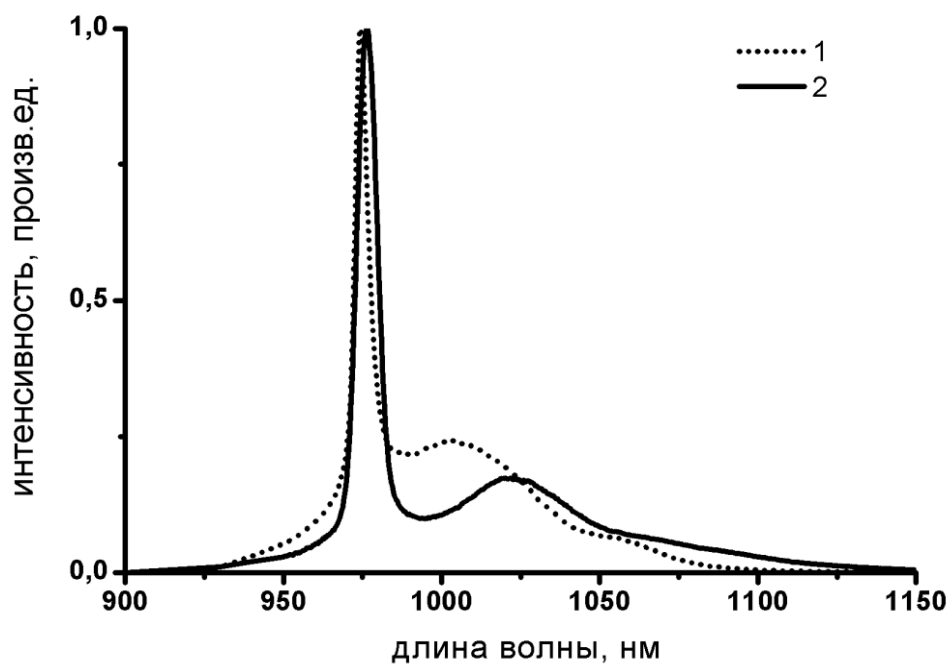

*Рис. 6. Спектры люминесценции ФС –1 и АС -2 световодов с примесью иттербия (нормированы на единицу)*

В то время как в большинстве измерений, проводимых в объемных образцах [8], [22] данное отношение равно 2.

### *Измерение времени жизни лазерного уровня*

Время жизни ионов $Yb^{3+}$ в возбужденном состоянии определялось по спаду люминесценции при возбуждении прямоугольными (со временем нарастания и спада ~1 мкс.) импульсами накачки с длиной волны ≈974 нм. Импульсы поступали в сердцевину отрезка иттербиевого световода длиной порядка нескольких сантиметров. Чтобы избежать влияния перепоглощения регистрация интенсивности люминесценции проводилась в направлении, перпендикулярном оси световода. Малая длина тестовых образцов и низкая мощность накачки (менее 1% от мощности насыщения) позволили проводить измерения в отсутствии генерации и с минимальной усиленной люминесценцией (расчетная величина усиления за проход менее 1 дБ).

Как известно, наблюдаемое время жизни не всегда совпадает с радиационным временем жизни из-за возможности безызлучательной релаксации. Но в нашем случае практически исключена возможность безызлучательных переходов из-за уникальности системы энергетических уровней $Yb^{3+}$. К тому же, обработка данных с цифрового осциллографа показала, что затухание люминесценции очень точно соответствует экспоненциальному закону (см. Рис. 7). Это еще один аргумент в пользу того, что наблюдалось действительно время жизни ионов $Yb^{3+}$ на уровне $^2\boldsymbol{F}_{5/2}$.

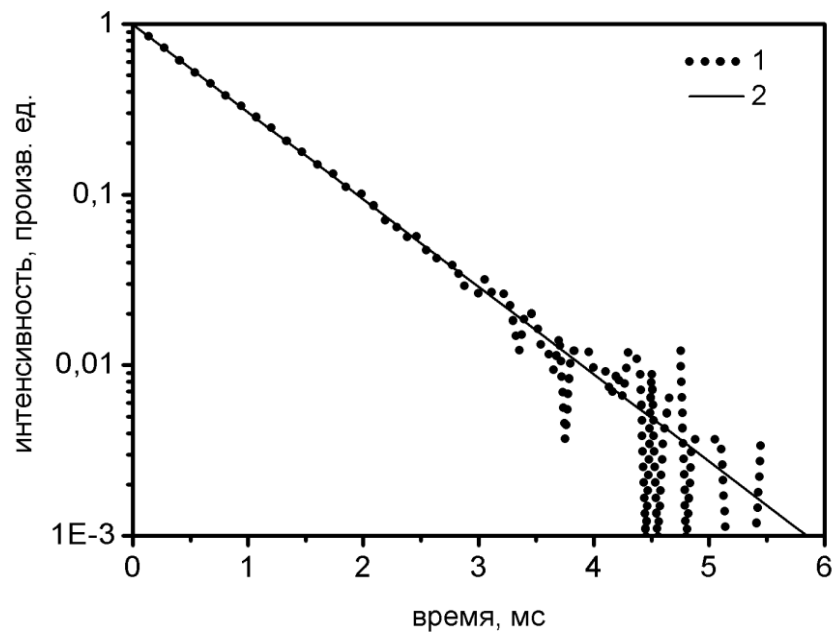

*Рис. 7 Затухание люминесценции в АС световоде от времени (1-экспериментальные данные, 2 – аппроксимация экспонентой).*

В исследованных световодах, времена жизни оказались равными $\tau$ = 1.45 мс в ФС матрице и $\tau$ = 0.83 мс в АС матрице. Причем, разброс значений времени релаксации для разных световодов (АС и ФС) одного типа (около 4%) лежит в пределах погрешности эксперимента. Для сравнения, в Таб. 1 приведены данные по времени жизни в иттербиевых световодах с различными легирующими добавками, опубликованные в работе [20].

Как видно, время жизни для алюминатной матрицы совпадает с высокой точностью, в то время как данные для фосфатной матрицы отличаются более чем на 10%. По-видимому, такое расхождение связано с отличием в составе фосфатного стекла. В пользу этого свидетельствует тот факт, что в указанной работе максимум люминесценции, соответствующий переходу с подуровня **e** на подуровень **b** (см. Рис. 4 б)) в фосфатном стекле находится в области 1018 нм, в то время как, в наших ФС световодах данный максимум люминесценции лежит вблизи 1012 нм. Указанный пик люминесценции в наших АС световодах находится на длине волны 1022 нм, что совпадает с данными из работы [20].

*Таб. 1 Время жизни уровня $^2F_{5/2}$ в различных стеклах.*

| Состав: плавленый кварц+ | $\tau$, мс [20] | $\tau$, мс Данная работа |
|---|---|---|
| Yb, P | 1.276 | 1,45 |
| Yb, Al | 0.830 | 0,83 |
| Yb, Ge | 0.760 | - |
| Yb | 0.755 | - |

### *Метод поглощения слабого сигнала.*

Коэффициент поглощения измерялся стандартным способом – т.н. методом "*облома*" – путем сравнения спектров пропускания двух различных длин световода. Свет от лампы накаливания вводился во вспомогательный одномодовый световод, к которому подваривался исследуемый образец. Противоположный конец исследуемого световода состыковывался с многомодовым световодом, соединенным со входом спектроанализатора. Затем образец укорачивался и измерения повторялись. Коэффициент поглощения определялся по разности двух полученных спектров. При такой методике ошибки связаны только с вариациями потерь на сварке исследуемого образца с многомодовым световодом. Но повторяемость таких соединений достаточно хороша, так как выходящая мода заведомо попадает в сердцевину многомодового световода.

В результате было установлено, что формы спектров поглощения для волоконных световодов одного типа (АС или ФС) практически не отличаются друг от друга. На Рис. 8 представлены спектры поглощения ФС и АС световодов.

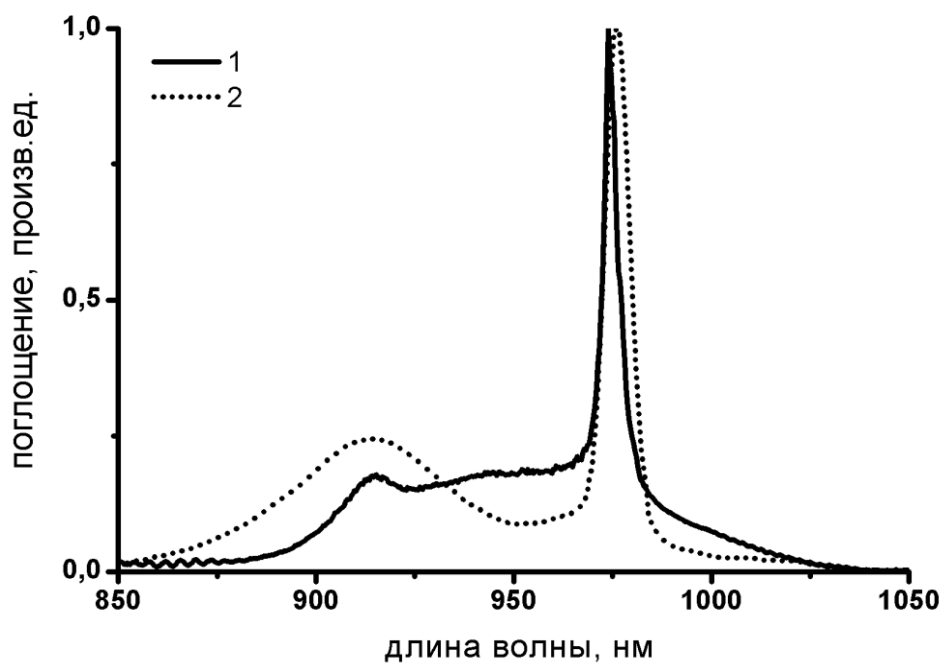

*Рис. 8. Спектры поглощения иттербиевых световодов: 1- ФС, 2- АС (нормированы на единицу)*

Максимум линии поглощения при комнатной температуре приходится на длину волны 976 нм в АС и на длину волны 974,5 нм в ФС световодах. Ширина центральных пиков поглощения на половине высоты соответственно 7,7 нм для АС и 4,7 нм для ФС световодов.

В Таб. 2 приведены данные, полученные для различных световодов: значения поглощения в сердцевине световодов - *α(974 нм)*, значения эффективной концентрации ионов Yb - $n_{eff}$ (по формуле (3)), сечения поглощения в максимуме линии - $\sigma_{abs}$ (по формуле (2)), а так же, значения интегрального сечения поглощения для рассматриваемой линии $\Sigma_{abs} = \int \sigma_{abs} d\lambda$, часто приводимые в литературе [7], [22], – с целью расширения возможностей для сравнения. Как уже отмечалось выше, измерения концентрации проводились по двум разным линиям ионов $Yb^{3+}$ - $M_\alpha$ и $L_\alpha$, поэтому приведены данные, полученные с использованием результатов для обеих линий. Измерения проводились при температуре 25°C.

*Таб. 2 Данные, полученные методом поглощения слабого сигнала*

| Световод | $\Delta n$ | $\alpha$(974 нм), дБ/мм | $n_{eff}(M_\alpha)$, $10^{20}$см$^{-3}$ | $n_{eff}(L_\alpha)$, $10^{20}$см$^{-3}$ | $\sigma_{abs}(M_\alpha)$, пм$^2$ | $\sigma_{abs}(L_\alpha)$, пм$^2$ | $\Sigma_{abs}*10^4$, пм$^3$ ($M_\alpha$) | $\Sigma_{abs}*10^4$, пм$^3$ ($L_\alpha$) |
|---|---|---|---|---|---|---|---|---|
| **ФС световоды** | | | | | | | | |
| Образец №1 | 0,012 | 1.3 | 3.0 | 2.1 | 1,0 | 1.4 | 2.1 | 3.0 |
| Образец №2 | 0,0139 | 0.76 | 1.9 | 1.4 | 0.9 | 1.3 | 1.9 | 2.7 |
| Образец №3 | 0,0166 | 2.4 | 5.0 | 3.6 | 1.1 | 1.5 | 2.3 | 3.3 |
| Образец №4 | 0,0164 | 1.7 | 3.9 | 2.8 | 1.0 | 1.4 | 2.1 | 3.0 |
| **АС световоды** | | | | | | | | |
| Световод | $\Delta n$ | $\alpha$(976 нм), дБ/мм | $n_{eff}(M_\alpha)$, $10^{20}$см$^{-3}$ | $n_{eff}(L_\alpha)$, $10^{20}$см$^{-3}$ | $\sigma_{abs}(M_\alpha)$, пм$^2$ | $\sigma_{abs}(L_\alpha)$, пм$^2$ | $\Sigma_{abs}*10^4$, пм$^3$ ($M_\alpha$) | $\Sigma_{abs}*10^4$, пм$^3$ ($L_\alpha$) |
| Образец №1 | 0.00815 | 0.70 | 1.3 | 0.6 | 1.3 | 2.7 | 3.4 | 7.1 |

Из Таб. 2 видно, что данные полученные методом рентгеноспектрального анализа по двум разным линиям иттербия существенно различаются. Причем, если для ФС световодов отличие составляет около 45%, то для АС световодов отличие более чем двукратное, что, по-видимому, обусловлено дополнительной линией алюминия в области 1,5 кэВ.

В работе [8] методом поглощения слабого сигнала были получены следующие значения сечения поглощения: $\sigma_{abs}$(peak)=1.2-1.5 пм$^2$ для фосфатных стекол ($\Sigma_{abs}$=3.6-4.1·$10^4$ пм$^3$), и $\sigma_{abs}$(peak)=1.5 пм$^2$ для алюминатного (0.5Yb$_2$O$_3$·65CaO·35Al$_2$O$_3$) стекла ($\Sigma_{abs}$=4·$10^4$ пм$^3$), причем состав стекол существенно отличался от состава стекол рассматриваемых в настоящей работе. Концентрация ионов иттербия в работе [8] определялась по плотности стекла. Видно, что сечения в фосфатных стеклах близки к нашим результатам для ФС световодов. Для алюминатных же стекол наблюдается существенное отличие полученных сечений от результатов по АС световодам. По-видимому, различия в полученных данных объясняются различием составов стекол, а так же плохой точностью измерений абсолютной концентрации иттербия в сетке стекла.

Окончательный вывод о том использование, какой линии $Yb^{3+}$ дает более достоверные результаты будет сделан после сравнения представленных в Таб. 2 данных с результатами измерений сечений другими методами.

### *Метод поглощение большого сигнала*

Схема установки по измерению сечения поглощения методом поглощения большого сигнала представлена на Рис. 9. В качестве накачки был использован неодимовый волоконный лазер, с излучением на длине волны 925 нм [21]. Длина волны излучения накачки была выбрана равной 925 нм потому, что сечение вынужденного излучения на этой длине волны практически равно нулю, а, следовательно, накачка не вызывает сброса инверсной населенности. Выход лазера накачки через мультиплексор соединен с исследуемым образцом. Были исследованы образцы длиной от 5 до 15 см. Не поглощенная в образце мощность $P_{out}$ измерялась непосредственно, а введенная мощность считалась пропорциональной мощности, выходящей из второго выхода мультиплексора $P_{ref}$. Отсутствие генерации в образце контролировалось с помощью спектроанализатора.

Основным недостатком данной схемы измерений следует считать, по-видимому, наличие сварки одного из выходных концов мультиплексора с исследуемым образцом. Величина потерь на указанной сварке существенно влияет на результат определения величины сечения поглощения.

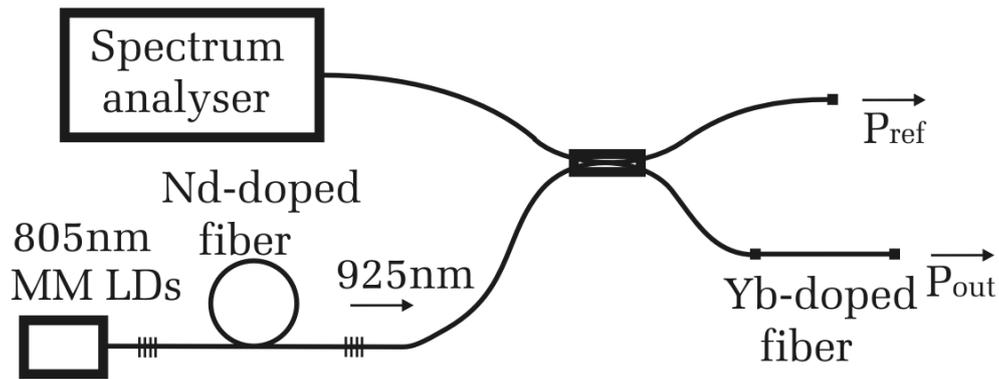

*Рис. 9. Экспериментальная установка для измерения поглощения большого сигнала*

Необходимо отметить, что измерить потери на данной сварке на нужной длине волны (в области поглощения иттербия) очень сложно, что существенно снижает точность результатов.

Мощность насыщения на длине волны 925 нм. в исследованных нами световодах составила порядка 10-20 мВт, поэтому необходимая для измерений мощность накачки не превышает 50 мВт.

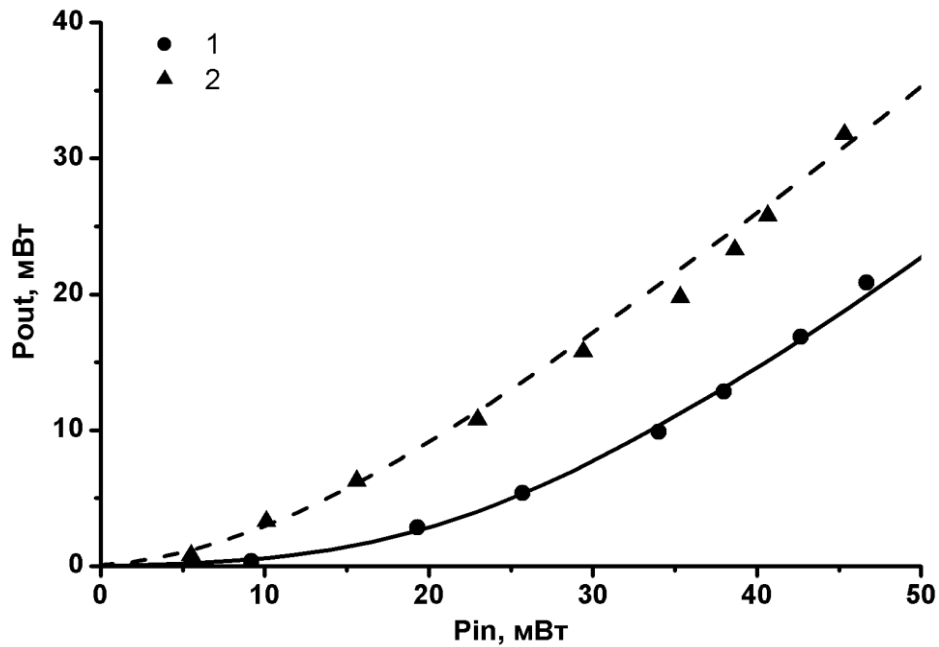

*Рис. 10. Зависимости мощности, выходящей из образца ФС световода-$P_{out}$ от входной мощности -$P_{in}$ для образцов различной длины:*
*1-108 мм, 2- 56 мм.*

Экспериментальные данные и расчетные зависимости непоглощенной мощности от введенной мощности для образца ФС световода представлены на Рис. 10. Метод поглощения большого сигнала позволяет определить сечение поглощения на определенной длине волны, совпадающей с длиной волны накачки. Нормируя весь спектр поглощения по данной точке можно определить сечение поглощения на любой длине волны в пределах данного спектра. Значения сечений поглощения на длине волны 925 нм и в максимуме, а так же интегральное сечение в ФС и АС световодах, полученные методом поглощения большого сигнала представлены в Таб. 3.

*Таб. 3 Сечения поглощения, полученные методом поглощения большого сигнала*

| Тип световода | $\sigma_{abs}(925nm)$, пм$^2$ | $\sigma_{abs}(peak)$, пм$^2$ | $\Sigma_{abs}*10^4$, пм$^3$ |
|---|---|---|---|
| ФС | 0.16 | 1.0 | 2.1 |
| АС | 0.52 | 2.3 | 6.1 |

Несмотря на то, что полученные данные согласуются с результатами, полученными в предыдущем параграфе для линии $M_\alpha$ (для ФС световода), по-видимому, значения сечений являются заниженными, так как при расчете не учитывались потери на сварке. Так, например, при учете потерь вносимых сваркой для АС световода (среднее значение около 0.3 дБ) значение сечения поглощения составит уже 2.7 пм$^2$. Данное обстоятельство свидетельствует о высокой чувствительности данного метода к потерям на сварке, что существенно снижает точность определения сечения поглощения.

### *Метод насыщения люминесценции*

В качестве источника накачки был выбран лазер на кристалле Ti:Al$_2$O$_3$ настроенный на длину волны излучения 948 нм. Такая длина волны удобна, так как сечение вынужденного излучения в данной области много меньше сечения поглощения. Излучение лазера вводилось в одномодовый световод, к которому приваривались исследуемые образцы длиной порядка одного сантиметра. Интенсивность люминесценции измерялась перпендикулярно оси тестируемого световода при помощи спектроанализатора. Мощность излучения измерялась непосредственно на выходном конце образца. Важно отметить, что интенсивность люминесценции необходимо фиксировать как можно ближе к выходному концу образца, чтобы выходная мощность соответствовала мощности, проходящей через измеряемое сечение. Для наблюдения насыщения люминесценции достаточно порядка 100 мВт входной мощности. Дальнейшее повышение мощности накачки практически не меняет мощность люминесценции. Экспериментальные данные и результаты численного расчета зависимости интенсивности люминесценции от проходящей мощности по формуле (15) представлены на Рис. 11. Сечения поглощения были найдены путем вариации величины $\sigma_a$ в формуле (15), так чтобы рассчитанная по этой формуле кривая совпадала с экспериментальными данными.

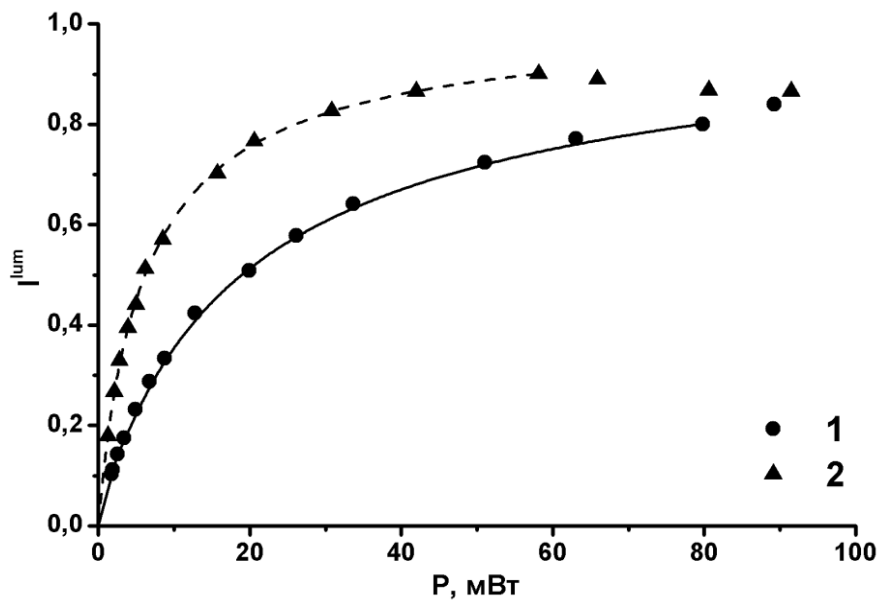

*Рис. 11. Интенсивность люминесценции от мощности накачки для разных типов световодов: 1-АС, 2-ФС*

Так же как и метод поглощения большого сигнала, данный метод позволяет определить сечение поглощения только на определенной длине волны, совпадающей с длиной волны накачки. Поэтому, и в этом случае для получения зависимости сечения поглощения от длины волны необходимо нормировать спектр поглощения по данной точке.

В Таб. 4 приведены значения сечений поглощения в ФС и АС световодах, легированных иттербием, полученные методом насыщения люминесценции.

*Таб. 4 Сечения поглощения, полученные методом насыщения люминесценции*

| Тип световода | $\sigma_{abs}(948нм)$, пм$^2$ | $\sigma_{abs}(peak)$, пм$^2$ | $\Sigma_{abs}*10^4$, пм$^3$ |
|---|---|---|---|
| ФС | 0.35 | 1.6 | 3.4 |
| АС | 0.47 | 2.8 | 7.4 |

Нам не известны работы, в которых данным методом было измерено сечение поглощения в стеклах с составом, близким нашему. Однако в работе [2] было измерено сечение поглощения в германосиликатных стеклах, легированных алюминием и бором. Средняя величина сечения поглощения в максимуме составила ~ 2.7 пм$^2$, при среднем времени жизни ~0.8 мс. При этом в зависимости от состава стекла, эти величины могли изменяться в пределах 30%. Для фосфоросиликатных стекол в работе [2] среднее время жизни составило 1,5 мс, данные о сечении отсутствуют. Таким образом, можно отметить совпадение времени жизни в фосфоросиликатной матрице в указанной работе со временем жизни в ФС световодах, измеренным в настоящей работе.

Метод насыщения люминесценции обладает существенным преимуществом по сравнению с методом поглощения большого сигнала в том, что он не чувствителен к потерям на сварках. Так же стоит отметить еще два преимущества: во-первых, в данном методе длина тестовых образцов значительно (до десяти раз) короче, чем в предыдущем методе, что сильно снижает вероятность вынужденных переходов из верхнего состояния в основное (суперлюминесценция, генерация), и, во-вторых, в расчеты не входит коэффициент поглощения в световоде. Таким образом, результаты, полученные данным методом, обладают более высокой степенью надежности по сравнению с результатами, полученными методом поглощения большого сигнала.

## *Метод определения сечения вынужденных переходов по спектрам люминесценции и времени жизни*

Для расчета сечения поглощения по сечению вынужденного излучения с использованием ТМК (см. формулы (18) и (19)) необходимо знать положение подуровней при расщеплении энергетических уровней $Yb^{3+}$ в кристаллическом поле. С этой целью нами были измерены спектры поглощения и люминесценции исследуемых типов иттербиевых световодов при температуре жидкого азота ($77^{o}K$), так как при низкой температуре уменьшается величина однородного уширения линии.

На Рис. 12 - Рис. 15 представлены эти результаты в логарифмическом масштабе. Приведенные экспериментальные спектры аппроксимировались линейной комбинацией гауссовых функций, каждая из которых соответствует переходу между определенными подуровнями. Поскольку измерения проводились при низкой температуре, то вследствие слабой заселенности верхних подуровней каждого уровня, можно утверждать, что спектры коэффициента поглощения образованы, главным образом, переходами с нижнего подуровня основного состояния на три подуровня возбужденного состояния, а спектры излучения образованы переходами с нижнего подуровня возбужденного состояния на каждый из четырех подуровней основного состояния. Формы уровней аппроксимировались с помощью распределения Гаусса, так как при такой низкой температуре основным фактором уширения линии является неоднородное уширение [18], вызванное неоднородностью матрицы стекла.

Очевидно, что спектры поглощения должны раскладываться на три гауссовых компоненты (по числу подуровней в возбужденном состоянии), в то время как спектры испускания на четыре компоненты (по числу подуровней в основном состоянии).

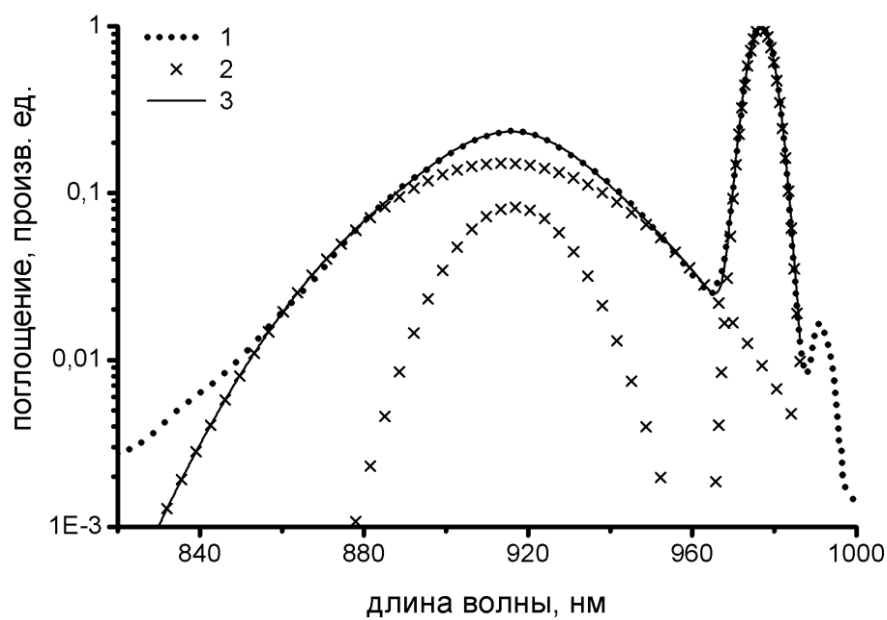

*Рис. 12 Коэффициент поглощения АС световода при температуре 77°K, нормированный на единицу. 1-экспериментальные данные, 2- разложение по гауссовым функциям, 3-сумма разложений по гауссовым функциям*

Как видно из Рис. 13 и Рис. 15 главный максимум (вблизи 975 нм) поглощения и люминесценции в ФС световодах раскладывается на как минимум две гауссовы компоненты. Причем интенсивность второй компоненты приблизительно в четыре раза ниже интенсивности основной компоненты. Такое поведение спектров может объясняться только присутствием дополнительной фазы в ФС стекле, т.е. в ФС световодах ионы $Yb^{3+}$ находятся в двух принципиально разных окружениях. Выделить другие максимумы в спектрах люминесценции и поглощения, относящиеся к дополнительной фазе стекла не удается вследствие их слабой интенсивности и большого неоднородного уширения.

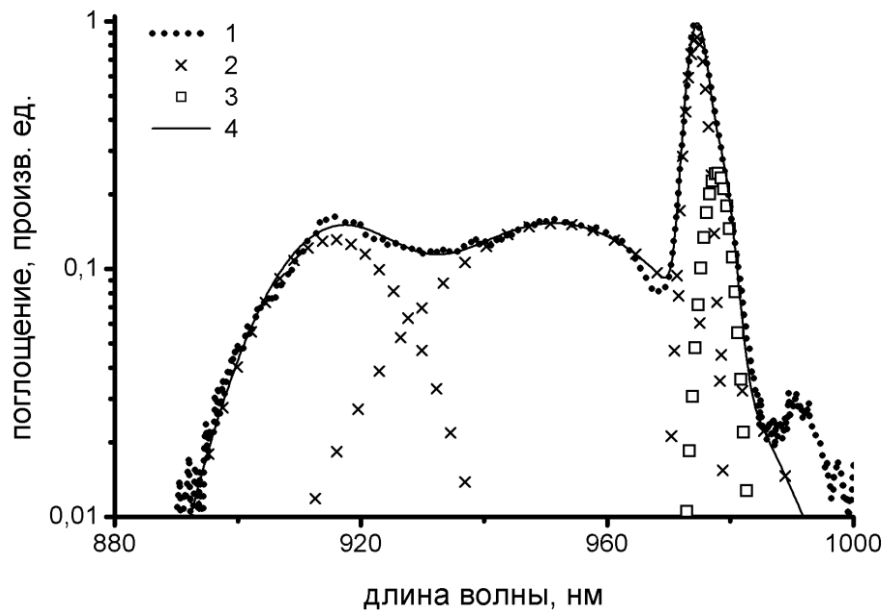

*Рис. 13 Коэффициент поглощения ФС световода при температуре 77°K, нормированный на единицу. 1-экспериментальные данные, 2- разложение по гауссовым функциям, 3-линия в разложении, указывающая на присутствие дополнительной фазы в фосфоросиликатном стекле, 4-сумма разложений по гауссовым функциям*

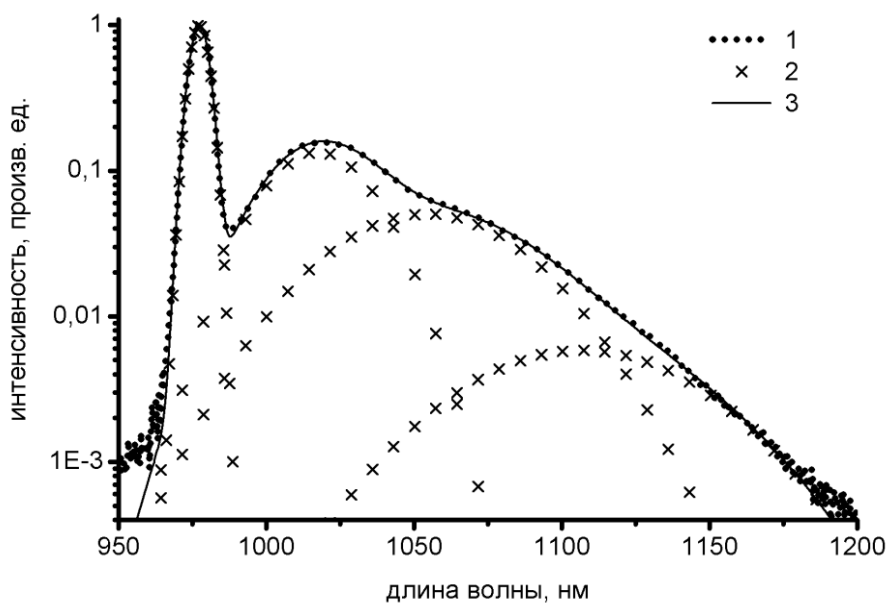

*Рис. 14 Спектр излучения АС световода при температуре 77°K. 1-экспериментальные данные, 2- разложение по гауссовым функциям, 3- сумма разложений по гауссовым функциям*

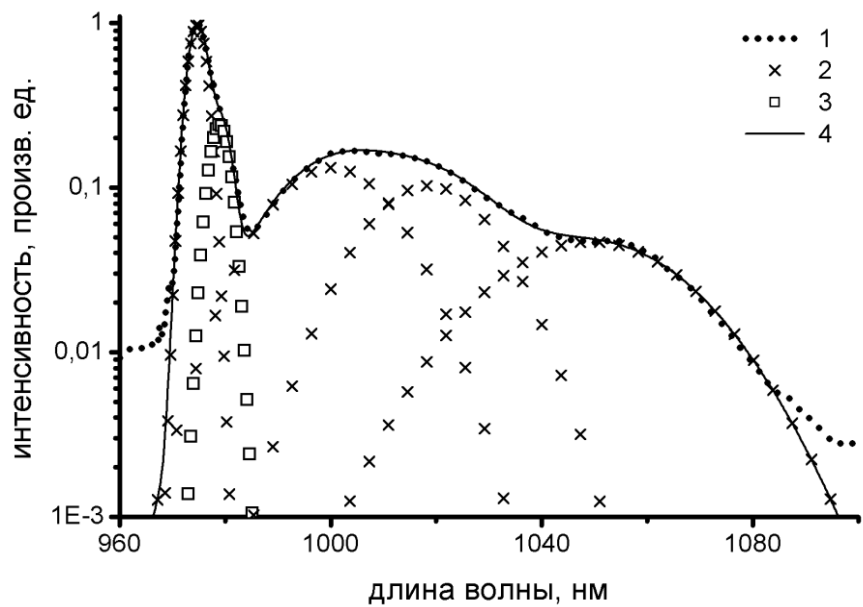

*Рис. 15 Спектр излучения ФС световода при температуре 77°K. 1-экспериментальные данные, 2- разложение по гауссовым функциям, 3-линия в разложении, указывающая на присутствие дополнительной фазы в ФС стекле, 4-сумма разложений по гауссовым функциям*

Приведенные разложения по гауссовым компонентам не являются единственно возможными, однако представляются наиболее вероятными. Отметим, что небольшие изменения в положения максимумов не вносят существенные изменения в результаты расчета. Так, например, в работе [13] для простоты предлагается считать, что расстояние между штарковскими компонентами одинаково. При этом утверждается, что сохраняется высокая точность результатов.

По спектрам поглощения и люминесценции, полученным при температуре 77°K, были найдены энергии расщепленных штарковских подуровней $Yb^{3+}$. Эти значения для разных матриц представлены в Таб. 5. Используя эти данные и формулу (19), стало возможным рассчитать сечение поглощения $\sigma_{abs}(\lambda)$ по известному сечению вынужденного излучения $\sigma_{emm}(\lambda)$ и наоборот.

*Таб. 5 Структура уровней Yb$^{3+}$ в АС и ФС матрицах стекла сердцевины волоконных световодов.*

| Уровень | Подуровень (см. Рис. 4) | Энергия, см$^{-1}$ | |
|---|---|---|---|
| | | **ФС световод** | **АС световод** |
| $^2F_{7/2}$ | a | 0 | 0 |
| | b | 260 | 400 |
| | c | 440 | 760 |
| | d | 740 | 1210 |
| $^2F_{5/2}$ | e | 10260 | 10245 |
| | f | 10520 | 10917 |
| | g | 10930 | 10940 |

Длины волн максимумов в спектрах сечения поглощения и вынужденного излучения практически (с точностью 0.5 нм) совпадают. Отношение сечения вынужденного излучения к сечению поглощения в максимуме $\kappa = \sigma_e/\sigma_a$ составило для ФС световодов $\kappa = 1.08$, для АС световодов $\kappa = 1.1$.

По времени жизни атома в возбужденном состоянии $\tau$ и спектрам люминесценции $I^{lum}(\lambda)$ были найдены абсолютные значения сечения эмиссии (см. формулу (21)). Исходя из полученных данных, было рассчитаны величины сечения поглощения.

Результаты расчетов сечений поглощения и люминесценции, а так же их отношение, рассчитанное при помощи ТМК вынесены в Таб. 6. Для удобства сравнения с опубликованными данными приведены значения интегрального сечения поглощения $\Sigma_{abs}$.

*Таб. 6 Сечения переходов в максимуме, их отношение, интегральное сечение поглощения.*

| Тип световода | $\sigma_{abs}$ (peak), пм$^2$ | $\sigma_{em}$ (*peak*), пм$^2$ | $\sigma_{em}/\sigma_{abs}$ (*peak*) | $\Sigma_{abs}*10^4$, пм$^3$ |
|---|---|---|---|---|
| ФС | 1.4 | 1.5 | 1.08 | 3.0 |
| АС | 2.7 | 3.0 | 1.1 | 7.1 |

В Таб. 7 приведены опубликованные данные, полученные по спектрам люминесценции и времени жизни.

*Таб. 7 Сечения поглощения $Yb^{3+}$ в различных стеклах. Опубликованные данные.*

| Источник | $\sigma_{abs}$ (peak), пм$^2$ | $\Sigma_{abs}$*10$^4$, пм$^3$ | Тип стекла |
|---|---|---|---|
| [7] | - | 3.6-4.0 | Фосфатное |
| [1] | 2.5 | - | Германосиликатное |
| [22] | - | 3.0-5.3 | Фосфатное |
| [15] | 1.5-2.6 | - | Фосфатное |
| [15] | 2.3 | - | Алюминатное |

При сравнении видно, что значения сечений поглощения в максимуме в фосфатных стеклах лежат в пределах 1.5-2.6 пм$^2$, что примерно соответствует интегральному сечению 3.0-5.6 пм$^3$. Таким образом, данные, полученные нами для ФС световодов, в пределах погрешности измерений согласуются с опубликованными результатами для фосфатных стекол. Так же можно отметить, что значение сечения поглощения в германосиликатном стекле [1] почти совпадает со значением сечения в наших АС световодах. Что касается существенного различия данных по сечению поглощения в алюминатном стекле, представленных в работе [15], по сравнению с данными для АС световодов, полученными в настоящей работе, то наиболее вероятной причиной такого отличия представляется большая разница в составах стекол.

Чтобы проверить корректность преобразования спектра излучения в спектр поглощения при помощи ТМК (см. формулы (18) и (19)), были проведены измерения спектра поглощения АС световодов в длинноволновой области спектра (1000-1200 нм.).

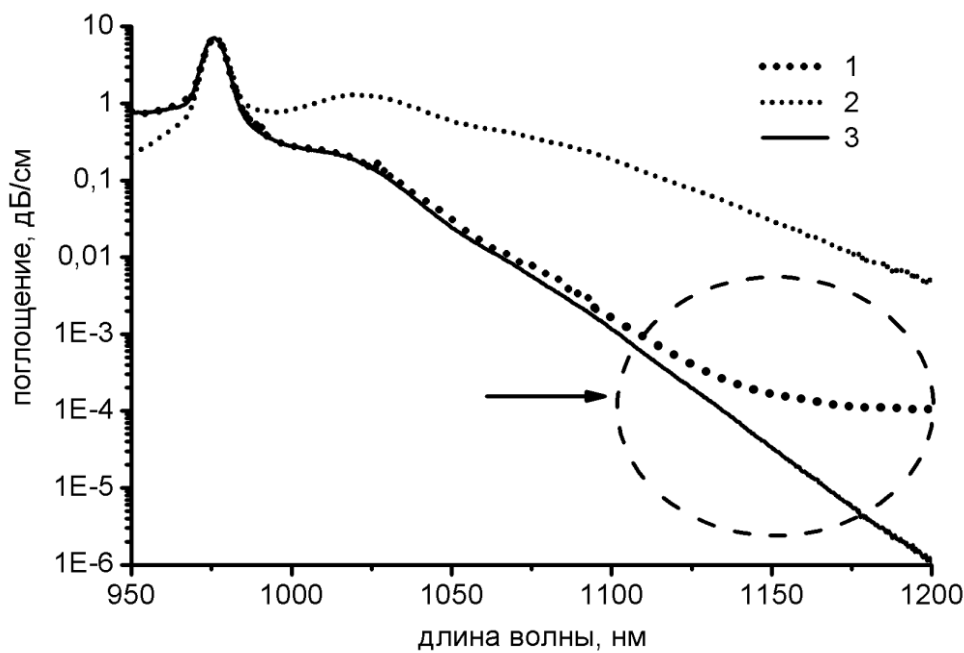

*Рис. 16 Сравнение формы спектров поглощения полученных экспериментально методом "облома" (кривая 1) и из спектра излучения при помощи ТМК (кривая 2- спектр излучения, нормированный в максимуме на пик поглощения, 3- спектр поглощения, полученный при помощи ТМК из кривой 2)*

На Рис. 16 представлены спектр поглощения АС световодов, полученный в результате прямых измерений (методом "*облома*"), и спектр поглощения, полученный с помощью ТМК из спектра излучения. Видно хорошее совпадение формы спектра, полученного с помощью ТМК с формой спектра, полученного экспериментально, в диапазоне 950-1100 нм. В диапазоне длин волн более 1100 нм наблюдается существенное различие в форме спектров (см. указатель на Рис. 16). Данное отличие объясняется наличием так называемых "серых" потерь в световоде, не связанных с поглощением иттербия. Так, в данном световоде величина серых потерь составляет около $10^{-4}$ (дБ/см). При учете серых потерь в диапазоне длин волн более 1100 нм разница между расчетным и экспериментальным спектрами существенно уменьшается. В целом, можно утверждать, что спектр поглощения, полученный из спектра излучения при помощи ТМК, совпадает со спектром поглощения, полученным в результате прямых измерений. Разница в величинах полученных сечений не превышает 10% в диапазоне 950-1030 нм. И составляет не более 25% вплоть до 1100нм. (с учетом серых потерь). При рассмотрении применимости ТМК было отмечено, что как однородное, так и неоднородное уширение может приводить к существенным искажениям формы спектра при его преобразовании. В данном случае, несмотря на различия полученных спектров, можно утверждать, что ТМК может быть использована для преобразования сечения вынужденного излучения в сечение поглощения.

## Анализ и обсуждение результатов

Величины максимальных значений сечений поглощения для всех методов измерений приведены в *Таб. 8*.

*Таб. 8 Сечения поглощения, измеренные различными методами*

| Тип световода | $\sigma_{abs}$(peak) пм² (метод) | | | | |
|---|---|---|---|---|---|
| | Поглощение слабого сигнала | | Поглощение большого сигнала | Насыщение люминесценции | По спектру люминесценции при помощи ТМК |
| | Линия $M_\alpha$ | Линия $L_\alpha$ | | | |
| Фосфатный | 1.0 | 1.4 | 1.0 | 1.5 | 1.4 |
| Алюминатный | 1.3 | 2.7 | 2.3 | 2.8 | 2.7 |

Полученные значения сечений поглощения лежат в довольно широком диапазоне. Однако, как легко заметить, результаты, полученные методом насыщения люминесценции, по спектру люминесценции при помощи ТМК и методом поглощения слабого сигнала с определением концентрации иттербия по линии $L_\alpha$ - отличаются не более чем на 10%, что укладывается в погрешность эксперимента. Как уже отмечалось выше, метод поглощения большого сигнала обладает принципиальным недостатком, а именно – невозможно корректно измерить потери на сварке тестируемого образца с мультиплексором. Наличие же таких потерь будет приводить к занижению величины сечения поглощения, что, по-видимому, и объясняет результат данного метода. Что касается сечений поглощения, полученных методом поглощения слабого сигнала с использованием рентгеноспектрального анализа по линии $M_\alpha$, то результаты в этом случае имеют, по-видимому, значительную систематическую ошибку.

Данные рассуждения приводят к выводу, что наиболее достоверными являются результаты, полученные при помощи ТМК по спектрам люминесценции и времени жизни, методом насыщения люминесценции и методом поглощения слабого сигнала с определением концентрации по линии $L_\alpha$ иттербия. Окончательное значение сечения поглощения было выбрано как среднее арифметическое по этим методам: $\sigma_{abs}(peak)=1.4$ пм$^2$ ($\Sigma_{abs}=3.1*10^4$ пм$^3$) – для ФС световодов, и $\sigma_{abs}(peak)=2.7$ пм$^2$ ($\Sigma_{abs}=7.1*10^4$ пм$^3$) –для АС световодов. Длины волн максимумов сечений вынужденного излучения и поглощения практически (с точностью 0.5 нм) совпадают для каждого из типов стекол и составляют 974.5 нм для фосфоросиликатного световода и 976 нм для алюмосиликатного световода при комнатной температуре. Соответствующие спектральные зависимости сечений поглощения и вынужденного излучения приведены в Приложении 2.

Значения сечения поглощения в ФС световодах согласуются с данными, по фосфатным стеклам представленными в литературе (см. Таб. 9).

*Таб. 9 Сечения поглощения в фосфатных стеклах. Опубликованные данные.*

| Ссылка | $\sigma_{abs}$ (peak), пм$^2$ | $\Sigma_{abs}*10^4$, пм$^3$ |
|---|---|---|
| [7] | - | 3.6-4.0 |
| [8] | 1.5 | 3.5-4.1 |
| [22] | - | 3.0-5.3 |
| [15] | 1.5-2.6 | - |

Что касается алюминатных стекол, то опубликованных данных по ним меньше. В работе [15] приводится величина сечения поглощения в максимуме $\sigma_{abs}(peak)=2.3$ пм$^2$ для стекла с составом 31.17Al$_2$O$_3$-47.01CaO-10.41MgO-10.41BaO-1Yb$_2$O$_3$. В работе [8] получены значения сечения поглощения в максимуме и интегральное сечение поглощения для состава XYb$_2$O$_3 \cdot$65CaO$\cdot$35Al$_2$O$_3$ - $\sigma_{abs}(peak)=1.4$ пм$^2$, $\Sigma_{abs}=4*10^4$ пм$^3$. Как видно, наши результаты отличаются от опубликованных данных довольно существенно. По-видимому, такое различие вызвано неодинаковостью состава стекол. С другой стороны, измерения сечений в германосиликатных световодах легированных алюминием и бором, выполненные в работах [1] и [2], дают значения сечений в пределах 2.5-2.7 пм$^2$, что весьма близко к полученным в настоящей работе значениям сечений в АС световодах с небольшой добавкой двуокиси германия.

На Рис. 17 и Рис. 18 представлены спектры сечений вынужденных переходов для ФС и АС световодов.

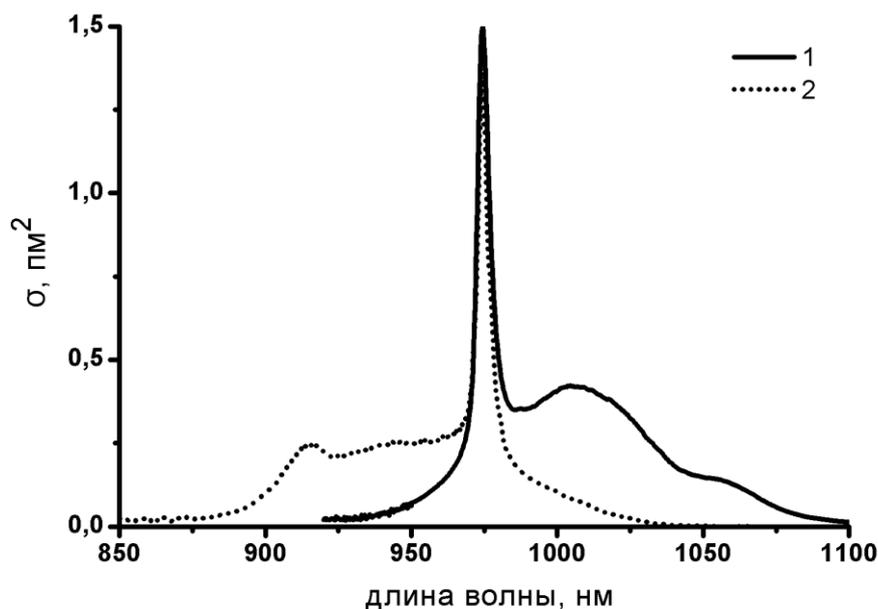

*Рис. 17 Сечения вынужденных переходов Yb$^{3+}$ в ФС волоконных световодах: 1–сечение вынужденного излучения, 2-сечение поглощения.*

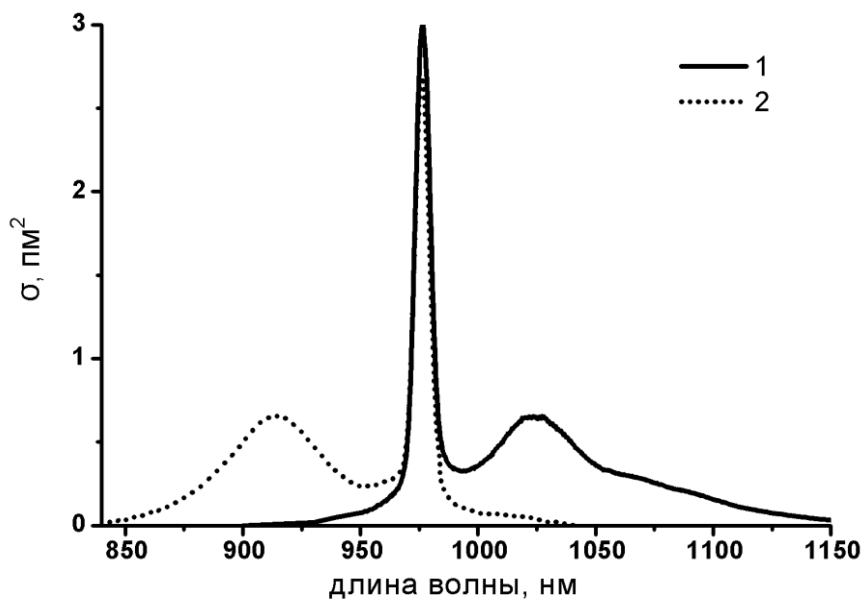

*Рис. 18 Сечения переходов Yb$^{3+}$ в АС волоконных световодах: 1-сечение вынужденного излучения, 2-сечение поглощения.*

Как видно из представленных графиков, сечения вынужденных переходов в фосфоросиликатном и алюмосиликатном стеклах отличаются не только по абсолютному значению сечения в максимуме (почти в два раза), но и качественно - по форме спектров. Причем, очень важным с точки зрения применения в волоконных лазерах является полная ширина по половине высоты (FWHM) основного максимума, которая в АС световодах составляет $\Delta\lambda_{FWHM}$=7.7 нм, что на 60% больше, нежели в ФС световодах $\Delta\lambda_{FWHM}$=4.8 нм. Такая разница приводит к гораздо большей толерантности лазера на основе АС световодов по отношению к длине волны излучения накачки, а значит к существенному снижению требований ее термостабилизации. Т.е. АС световоды с этой точки зрения выглядят более перспективными. С другой стороны, в АС световодах наблюдается провал в спектре поглощения в области 930-970 нм, что делает ФС световоды более удобными для использования в этой области, так как у них в этой области наблюдается плато в спектре поглощения.

На Рис. 19 представлены нормированные в максимуме на единицу спектры сечений поглощения и вынужденного излучения в рассматриваемых стеклах в длинноволновой области. Подобная информация может быть весьма полезной при моделировании лазеров и усилителей. Спектры сечения поглощения получены из спектров люминесценции при помощи ТМК. На врезке представлены нормированные на единицу в максимуме сечения поглощения вблизи главного пика.

Для сравнения на Рис. 19 так же представлены сечения поглощения и люминесценции $Yb^{3+}$ в германосиликатном стекле, легированном алюминием и бором [2]. Абсолютное значение сечения поглощения и люминесценции в указанной работе составило 2,7 пм$^2$.

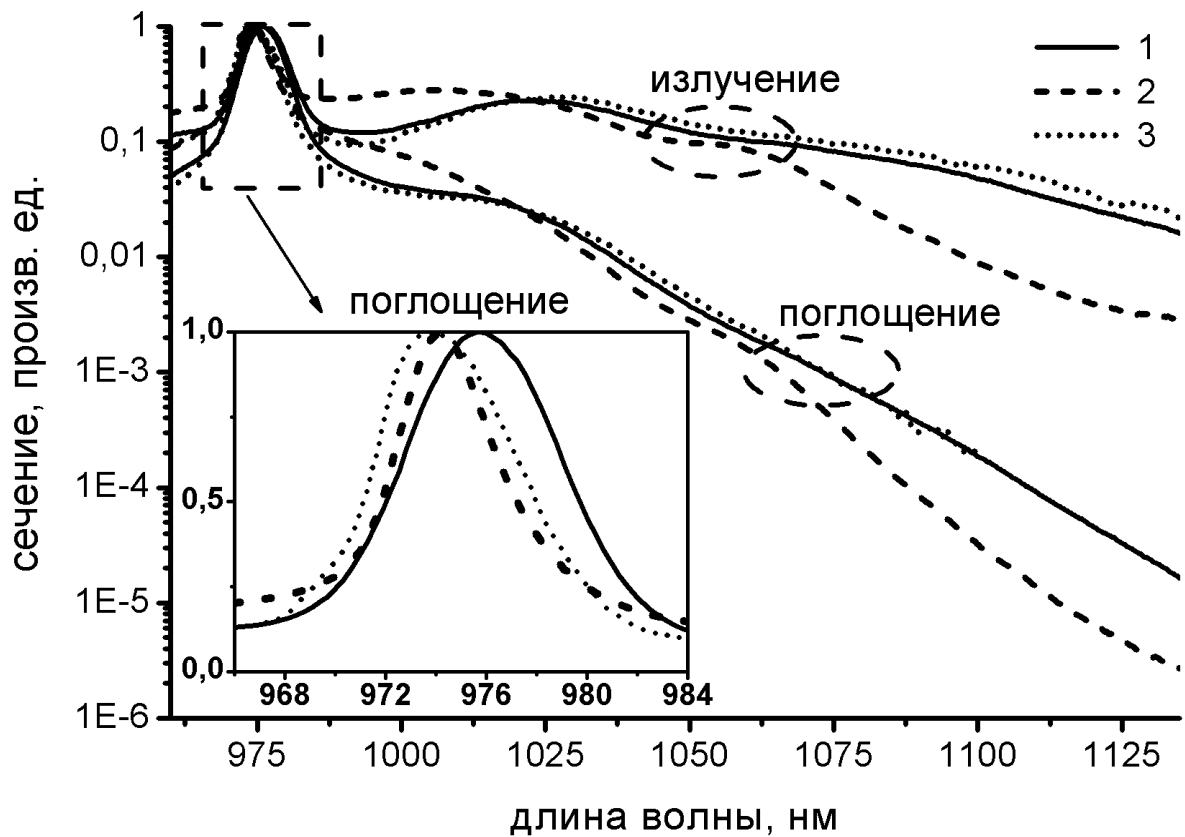

*Рис. 19 Сечения поглощения и вынужденного излучения (нормированные в максимуме на единицу) ионов $Yb^{3+}$ в длинноволновой области для АС световодов (1) и ФС световодов (2), 3-спектры германосиликатного стекла, допированного алюминием и бором [2].*

## Заключение

В настоящей работе измерены времена жизни ионов $Yb^{3+}$ на верхнем лазерном уровне в алюмосиликатных и фосфоросиликатных световодах. Определено положение подуровней в мультиплетах лазерных энергетических уровней $Yb^{3+}$ в указанных матрицах. Несколькими различными методами определены сечения вынужденных переходов $^2F_{5/2} \to {}^2F_{7/2}$ ионов $Yb^{3+}$ как функции длины волны, причем методы поглощения слабого сигнала, насыщения люминесценции и метод определения сечения вынужденного излучения по спектру люминесценции и времени жизни дали согласующиеся между собой значения сечений. Продемонстрировано, что использование теории МакКамбера [11] для расчета сечения поглощения по сечению вынужденного излучения в рассматриваемой системе дает удовлетворительные результаты.

Полученные данные могут быть использованы для численного моделирования волоконных лазеров и усилителей. В частности, значительное снижение сечения вынужденного излучения ионов $Yb^{3+}$ в фосфоросиликатной матрице в спектральной области $\lambda \geq 1080$ нм., ограничивает возможности использования лазеров на световодах такого типа в этой области по сравнению с лазерами на алюмосиликатных световодах.

# Приложение 1. Вывод сечения вынужденного излучения через коэффициенты Эйнштейна.

Рассмотрим простую двухуровневую систему, находящуюся в тепловом равновесии с излучением. В таких условиях количество вынужденных переходов снизу вверх, сверху вниз и количество спонтанных переходов в единицу времени связаны соотношением:

$$N_1 \rho(v) B_{12} = N_2 \rho(v) B_{21} + N_2 A_{21} \qquad (23)$$

где $\rho(v)$ - спектральная плотность мощности излучения на частоте $v$, а $B_{12}$, $B_{21}$ и $A_{21}$ – коэффициенты Эйнштейна.

Из (23) определяется равновесная спектральная плотность энергии фотонов:

$$\rho(v) = \frac{A_{21}/B_{21}}{B_{12} N_1 / B_{21} N_2 - 1} \qquad (24)$$

Отношение населенностей на верхнем и нижнем уровнях описывается формулой Больцмана (если уровни не вырождены):

$$\frac{N_2}{N_1} = \exp(-hv/kT) \qquad (25)$$

Подставляя выражение (25) в формулу (24), получим

$$\rho(v) = \frac{A_{21}/B_{21}}{B_{12}/B_{21} \exp(-hv/kT) - 1} \qquad (26)$$

Согласно формуле Планка для излучения черного тела:

$$\rho(v) = \frac{8\pi h v^3 n^3}{c^3 [\exp(hv/kT) - 1]} \qquad (27)$$

где n – показатель преломления. Сравнивая (27) и (26), найдем, что данные выражения совпадают при любой температуре, если:

$$B_{12} = B_{21} = B \qquad (28)$$

$$\frac{A}{B} = \frac{8\pi h \nu^3 n^3}{c^3} \qquad (29)$$

Из (27) и (29) следует, что

$$\frac{A}{B\rho(\nu)} = \exp\left[\frac{h\nu}{kT}\right] - 1 \qquad (30)$$

Слева здесь стоит отношение вероятностей спонтанных и вынужденных переходов с уровня 2 на уровень 1. Из (30) следует, что при тепловом равновесии для $\nu \ll \frac{kT}{h}$ преобладают вынужденные переходы, а при $\nu \gg \frac{kT}{h}$ – спонтанные. Равенство вероятностей спонтанных и вынужденных переходов имеет место при $\exp\left[\frac{h\nu}{kT}\right] - 1 = 1$. Учитывая, что $\left\{\exp\left[\frac{h\nu}{kT}\right] - 1\right\}^{-1}$ равно среднему числу фотонов в моде, то можно сказать, что вероятности спонтанного и вынужденного переходов равны при спектральной плотности излучения, соответствующему одному фотону на моду.

Спектральная плотность числа мод на единицу объема равна

$$\frac{8\pi n^3 \nu^2}{c^3}$$

Если на каждую моду приходится один квант излучения, то это число равно спектральной плотности числа квантов в единице объема. Отсюда следует, что спектральная плотность потока квантов, падающих на один атом, равна

$$\Phi = \frac{8\pi n^3 \nu^2}{c^3} \cdot \frac{c}{n} = \frac{8\pi n^2 \nu^2}{c^2}$$

и вероятность вынужденных переходов в диапазоне $d\nu$ под действием этого потока на атом, который может взаимодействовать с излучением на различных частотах с сечением вынужденного перехода $\sigma(\nu)$ есть

$$\Phi\sigma(\nu)d\nu = \frac{8\pi n^2 \nu^2}{c^2}\sigma(\nu)d\nu = \frac{P_{lum}(\nu)}{h\nu}d\nu \qquad (31)$$

где $P_{lum}(\nu)$ – спектральная плотность мощности излучения люминесценции. Переходя в соотношении (31) к длинам волн, получим

$$\Phi\sigma(\lambda)d\lambda = \frac{8\pi n^2 c}{\lambda^4}\sigma(\lambda)d\lambda = \frac{P_{lum}(\lambda)\lambda}{hc}d\lambda \qquad (32)$$

Из последнего соотношения сразу следует, что $\sigma(\lambda) \sim \lambda^5 \cdot P_{lum}(\lambda)$.

Но суммарная по всем частотам вероятность перехода равна вероятности спонтанного распада уровня (при выбранной плотности мощности излучения), поэтому интегрируя (32), получим

$$\frac{1}{\tau} = 8\pi n^2 c \int \frac{\sigma(\lambda)}{\lambda^4}d\lambda = \int \frac{P_{lum}(\lambda)\cdot\lambda}{hc}d\lambda \qquad (33)$$

что совпадает с соотношением (20).

# Приложение 2. Сечения поглощения и вынужденного излучения Yb$^{3+}$ в АС и ФС световодах.

| λ, нм | АС | | ФС | | λ, нм | АС | | ФС | |
|---|---|---|---|---|---|---|---|---|---|
| | $\sigma_{emm}$, пм$^2$ | $\sigma_{abs}$, пм$^2$ | $\sigma_{emm}$, пм$^2$ | $\sigma_{abs}$, пм$^2$ | | $\sigma_{emm}$, пм$^2$ | $\sigma_{abs}$, пм$^2$ | $\sigma_{emm}$, пм$^2$ | $\sigma_{abs}$, пм$^2$ |
| 848 | 2,2E-5 | 0,033 | 1,0E-5 | 0,0011 | 988 | 0,36 | 0,18 | 0,41 | 0,17 |
| 852 | 3,5E-5 | 0,041 | 1,9E-5 | 0,0015 | 992 | 0,33 | 0,14 | 0,42 | 0,14 |
| 856 | 6,3E-5 | 0,057 | 2,0E-5 | 0,0035 | 996 | 0,33 | 0,11 | 0,45 | 0,12 |
| 860 | 1,1E-4 | 0,075 | 2,2E-5 | 0,0057 | 1000 | 0,36 | 0,099 | 0,47 | 0,11 |
| 864 | 1,7E-4 | 0,090 | 2,8E-5 | 0,0090 | 1004 | 0,40 | 0,092 | 0,49 | 0,091 |
| 868 | 2,7E-4 | 0,11 | 4,6E-5 | 0,014 | 1008 | 0,46 | 0,088 | 0,49 | 0,075 |
| 872 | 4,4E-4 | 0,14 | 7,0E-5 | 0,017 | 1012 | 0,53 | 0,084 | 0,48 | 0,061 |
| 876 | 6,9E-4 | 0,17 | 1,0E-4 | 0,023 | 1016 | 0,60 | 0,078 | 0,45 | 0,048 |
| 880 | 0,0011 | 0,21 | 1,6E-4 | 0,028 | 1020 | 0,65 | 0,070 | 0,43 | 0,037 |
| 884 | 0,0017 | 0,26 | 2,4E-4 | 0,033 | 1024 | 0,65 | 0,059 | 0,39 | 0,029 |
| 888 | 0,0026 | 0,31 | 3,8E-4 | 0,042 | 1028 | 0,65 | 0,049 | 0,35 | 0,021 |
| 892 | 0,0039 | 0,37 | 6,2E-4 | 0,054 | 1032 | 0,60 | 0,038 | 0,30 | 0,015 |
| 896 | 0,0058 | 0,43 | 0,0011 | 0,072 | 1036 | 0,55 | 0,029 | 0,26 | 0,011 |
| 900 | 0,0086 | 0,50 | 0,0019 | 0,10 | 1040 | 0,49 | 0,022 | 0,21 | 0,0076 |
| 904 | 0,012 | 0,57 | 0,0032 | 0,14 | 1044 | 0,44 | 0,016 | 0,19 | 0,0057 |
| 908 | 0,017 | 0,62 | 0,0056 | 0,19 | 1048 | 0,39 | 0,012 | 0,18 | 0,0044 |
| 912 | 0,022 | 0,65 | 0,0085 | 0,23 | 1052 | 0,35 | 0,0090 | 0,17 | 0,0036 |
| 916 | 0,029 | 0,65 | 0,012 | 0,24 | 1056 | 0,33 | 0,0072 | 0,16 | 0,0029 |
| 920 | 0,034 | 0,62 | 0,013 | 0,22 | 1060 | 0,31 | 0,0057 | 0,15 | 0,0022 |
| 924 | 0,039 | 0,57 | 0,016 | 0,21 | 1064 | 0,30 | 0,0046 | 0,13 | 0,0016 |
| 928 | 0,044 | 0,51 | 0,021 | 0,22 | 1068 | 0,29 | 0,0038 | 0,11 | 0,0011 |
| 932 | 0,048 | 0,44 | 0,026 | 0,23 | 1072 | 0,27 | 0,0030 | 0,083 | 7,4E-4 |
| 936 | 0,050 | 0,38 | 0,033 | 0,23 | 1076 | 0,26 | 0,0024 | 0,064 | 4,9E-4 |
| 940 | 0,053 | 0,32 | 0,044 | 0,25 | 1080 | 0,23 | 0,0018 | 0,049 | 3,1E-4 |
| 944 | 0,057 | 0,28 | 0,057 | 0,26 | 1084 | 0,22 | 0,0015 | 0,038 | 2,1E-4 |
| 948 | 0,062 | 0,24 | 0,069 | 0,26 | 1088 | 0,21 | 0,0012 | 0,030 | 1,4E-4 |
| 952 | 0,074 | 0,23 | 0,090 | 0,26 | 1092 | 0,19 | 9,5E-4 | 0,025 | 9,7E-5 |
| 956 | 0,095 | 0,24 | 0,11 | 0,26 | 1096 | 0,18 | 7,3E-4 | 0,019 | 6,5E-5 |
| 960 | 0,13 | 0,26 | 0,14 | 0,26 | 1100 | 0,16 | 5,6E-4 | 0,016 | 4,5E-5 |
| 964 | 0,17 | 0,28 | 0,18 | 0,27 | 1104 | 0,14 | 4,2E-4 | 0,013 | 3,2E-5 |
| 968 | 0,26 | 0,35 | 0,25 | 0,32 | 1108 | 0,12 | 3,2E-4 | 0,011 | 2,3E-5 |
| 969 | 0,34 | 0,44 | 0,28 | 0,34 | 1112 | 0,11 | 2,4E-4 | 0,0093 | 1,7E-5 |
| 970 | 0,46 | 0,57 | 0,34 | 0,39 | 1116 | 0,098 | 1,9E-4 | 0,0078 | 1,2E-5 |
| 971 | 0,70 | 0,83 | 0,45 | 0,50 | 1120 | 0,088 | 1,4E-4 | 0,0068 | 9,0E-6 |
| 972 | 1,08 | 1,21 | 0,73 | 0,76 | 1124 | 0,076 | 1,1E-4 | 0,0062 | 7,0E-6 |
| 973 | 1,58 | 1,68 | 1,16 | 1,15 | 1128 | 0,071 | 8,5E-5 | 0,0057 | 5,5E-6 |
| 974 | 2,14 | 2,17 | 1,46 | 1,38 | 1132 | 0,061 | 6,3E-5 | 0,0052 | 4,4E-6 |
| 975 | 2,65 | 2,55 | 1,43 | 1,29 | 1136 | 0,055 | 4,9E-5 | 0,0051 | 3,6E-6 |
| 976 | 2,97 | 2,69 | 1,18 | 1,01 | 1140 | 0,047 | 3,6E-5 | 0,0049 | 3,0E-6 |
| 977 | 2,94 | 2,53 | 0,93 | 0,75 | 1144 | 0,042 | 2,8E-5 | 0,0044 | 2,4E-6 |
| 978 | 2,71 | 2,22 | 0,73 | 0,56 | 1148 | 0,035 | 2,0E-5 | 0,0042 | 2,0E-6 |
| 979 | 2,28 | 1,77 | 0,59 | 0,43 | 1152 | 0,031 | 1,6E-5 | 0,0041 | 1,7E-6 |
| 980 | 1,78 | 1,32 | 0,50 | 0,35 | 1156 | 0,027 | 1,1E-5 | 0,0039 | 1,3E-6 |
| 981 | 1,29 | 0,91 | 0,43 | 0,29 | 1160 | 0,023 | 8,6E-6 | 0,0036 | 1,1E-6 |
| 982 | 0,91 | 0,61 | 0,40 | 0,25 | 1164 | 0,021 | 6,8E-6 | 0,0035 | 9,1E-7 |
| 983 | 0,67 | 0,43 | 0,37 | 0,22 | 1168 | 0,018 | 4,9E-6 | 0,0033 | 7,5E-7 |
| 984 | 0,53 | 0,32 | 0,36 | 0,21 | 1172 | 0,014 | 3,5E-6 | 0,0030 | 5,9E-7 |
| 985 | 0,45 | 0,26 | 0,35 | 0,19 | 1176 | 0,014 | 3,1E-6 | 0,0027 | 4,6E-7 |
| 986 | 0,41 | 0,23 | 0,35 | 0,18 | 1180 | 0,012 | 2,2E-6 | 0,0026 | 3,9E-7 |

# Литература